\documentclass[aps,10pt,twocolumn,amsmath,amssymb,prx,longbibliography]{revtex4-2}
\usepackage{graphicx}
\usepackage{color}

\usepackage{bm}

\usepackage{hyperref}

\begin{document}

\title{Renormalization approach to the analysis and design\\ of Hermitian and non-Hermitian interfaces}

\author{Henning Schomerus}
\affiliation{Department of Physics, Lancaster University, Lancaster, LA1 4YB, United Kingdom}
\begin{abstract}
I describe a concrete and efficient real-space renormalization approach that provides a unifying perspective on interface states in a wide class of Hermitian and non-Hermitian models, irrespective of whether they obey a traditional bulk-boundary principle or not.
The emerging interface physics are governed by a flow of microscopic interface parameters, and the properties of interface states become linked to the fixed-point topology of this flow. In particular, the quantization condition of interface states converts identically into the question of the convergence to unstable fixed points.
As its key merit, the approach can be directly applied to concrete models and utilized to design interfaces that induce states with desired properties, such as states with a  predetermined and possibly symmetry-breaking energy. I develop the approach in general, and then demonstrate these features in various settings, including for the design of circular, triangular and square-shaped complex dispersion bands and associated arcs at the edge of a two-dimensional system. Furthermore, I describe how this approach transfers to nonlinear settings, and demonstrate the efficiency, practicability and consistency of this extension for a paradigmatic model of topological mode selection by distributed saturable gain and loss.
\end{abstract}

\maketitle

\section{Introduction}
\label{sec:intro}
States pinned to interfaces are central to our understanding of a wide range of quantum and classical wave systems. For instance, these states are the most directly visible signature of topological phases, where they induce unique physical phenomena such as directed transport, charge fractionalization, and anomalous response to external perturbations \cite{Has10,Qi11,Beenakker2015}.
In Hermitian systems, the appearance of symmetry-protected states is governed by the bulk-edge correspondence \cite{Asboth2016}, but
this breaks down when the interfaces violate the relevant bulk symmetries, which is frequently unavoidable due to the nature of these symmetries \cite{Shiozaki14} or realistic physical boundary effects \cite{kittel2018introduction}. This blurs their distinction from ubiquitous and equally significant conventional defect states, which appear by mechanisms that cannot be described in topological frameworks.
An even richer phenomenology emerges when the underlying system is described by a non-Hermitian effective Hamiltonian, as one encounters in a large assortment of recently explored settings and applications \cite{moiseyev2011non,Kaw19,Oza19,zhou2019,ashida2020,Ota20,Okuma2022,Price2022}. Such systems can display complex edge dispersions forming bands and arcs \cite{Yang2017,Ni2018,Mal18,Ber19}, as well as phase transitions where interface states appear that do not have a Hermitian counterpart \cite{Mal15,Lee2016,Lang18,Mostafavi2020}, while the non-Hermitian skin effect can lead them to support a macroscopic number of bulk states distorted towards an edge \cite{Yao18,Kun18,Lee19,Imu19,Yokomizo19,Borgnia2020}.
However, the understanding of interface states in these systems is hampered by a manifest break-down of the bulk-boundary correspondence, and often relies on detailed studies of abstract model-specific features.

Here, I develop a unifying description of interface states based on a practically useful and simple real-space renormalization approach, which can be directly applied to both analyze and design such states in a wide range of concrete and paradigmatic Hermitian and non-Hermitian model systems, irrespective of whether they obey a traditional bulk-boundary principle or not.
The approach links the interface states to nontrivial fixed points in a dynamical flow, which directly occurs in a low-dimensional space of interface parameters of the microscopic model. In particular, the quantization condition of the interface states is met precisely for energies and system parameters that allow convergence to an unstable fixed point.
This connects the study of interface states to the rich phenomenology and powerful toolset of dynamical systems theory, resulting in conceptual and concrete insights of a very different nature than those afforded by other general methods, such as calculations based on transfer matrices \cite{dwivedi2018,Kun19}.
In particular, the presented approach can be directly employed to efficiently design interfaces that support states with desired properties.

This connection is usefully developed in a concrete, motivating example.
This is carried out in Sec.~\ref{sec:ssh}, where we equip a Su-Schrieffer Heeger chain (dimer chain with alternating bulk couplings $s$ and $t$ \cite{Su79}) with an analytically determined symmetry-breaking edge potential $u$ that induces a state of a predetermined, non-vanishing energy $E$ (see Figs.~\ref{fig:ssh} and ~\ref{fig:ssh2}).
This potential follows from the nontrivial fixed point of a simple one-parameter map. As long as $|E|<||s|-|t||$ is in the bulk gap, the construction works irrespective of the relative size of the couplings, hence, both in the topologically trivial ($|s|>|t|$) and  nontrivial ($|s|<|t|$) phase of the bulk.

\begin{figure*}[t]
\includegraphics[width=0.8\linewidth]{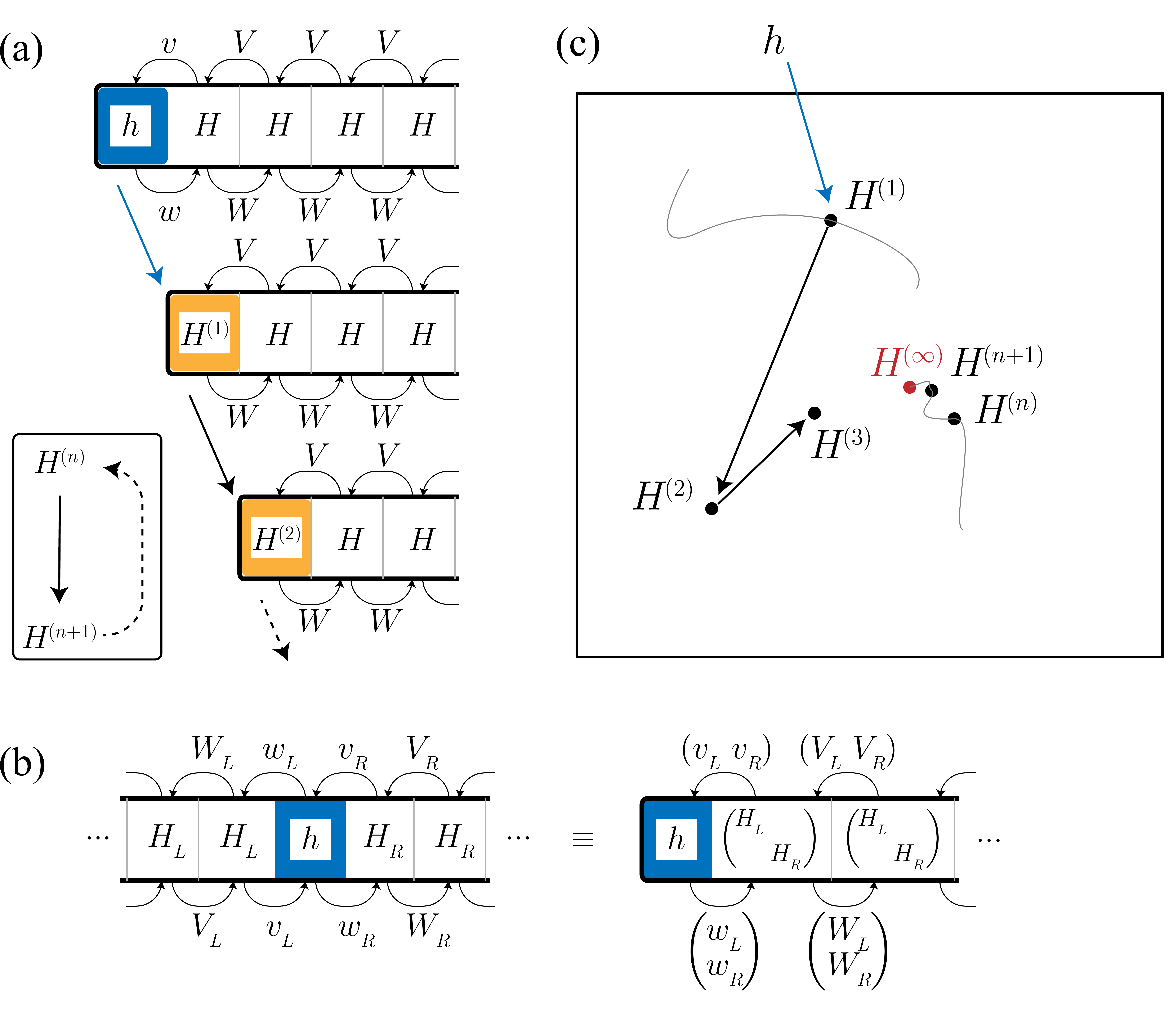}
\caption{General interface renormalization approach. (a) Renormalization structure at an edge, with local Hamiltonian $h$ coupled to the bulk by matrices $v$ and $w$, and the bulk represented by analogous matrices $H$, $V$ and $W$. The renormalization procedure
results in a sequence of renormalized edge Hamiltonians $H^{(n)}$, which are obtained via the map \eqref{eq:genmap}.
The procedure also applies to interfaces between two bulk regions (subscripts $L$ and $R$), which can be transformed into an edge configuration as shown in panel (b).
(c) Graphical illustration of the resulting map in the effective interface parameter space. The bare interface parameters provide the initial conditions. As shown in Sec.~\ref{sec:stabquant}, normalizable interface states appear at energies and system parameters where the map converges to fixed points with unstable directions, which requires the initial conditions to sit on their stable manifold (sketched as a thin gray curve). This provides a universal and efficient description of these states, irrespective of whether a conventional bulk-boundary principle applies or not.
}
\label{fig:generalapproach}
\end{figure*}

The general approach is then presented in Sec.~\ref{sec:general}. In it, bulk data determines a renormalization map of the interface parameters (see Fig.~\ref{fig:generalapproach}), and interface states arise for convergence to nontrivial fixed points, which are precisely those with unstable directions. This results in a clear separation of bulk and boundary data, irrespective whether the system obeys a conventional bulk-boundary principle or not, and makes the approach suitable both for the efficient analysis and design of interface states (see Fig.~\ref{fig:generalapproach2}).

In the remainder of the work, these general renormalization principles are illustrated in further examples, which demonstrate how the approach offers a compact and often analytic understanding of interface states in Hermitian and non-Hermitian systems, and serves as an efficient tool to design their concrete features.
Section \ref{sec:dimer} describes an interface between two bulk regions in a general dimer chain (Fig.~\ref{fig:dimer}), and illustrates in detail the fixed point manifolds, role of bulk symmetries, non-Hermitian phase transitions, and interface state design  (Figs.~\ref{fig:dimer}, \ref{fig:flow}, and \ref{fig:dimerdesign}).
Section \ref{sec:ladder} discusses a non-Hermitian nonreciprocal two-legged ladder, which supports interface states without a Hermitian counterpart (Figs.~\ref{fig:mpsconvergence} and \ref{fig:mpsdesign}).
Section \ref{sec:bandsandarcs} details the design of complex edge dispersions in a two-dimensional system, resulting in bands that approximate circles, triangles, or squares in the complex energy plane, and support unidirectional transport (Figs.~\ref{fig:sshq1dsketch} and \ref{fig:sshq1dall}). Section \ref{sec:nonlin} describes the extension of the approach to nonlinear models, and demonstrates the practicability and efficiency of this extension for a dimer chain with distributed saturable gain and loss, which delivers original insights into topological mode selection (see Fig.~\ref{fig:nonlin}).

The conclusions in Sec.~\ref{sec:conclusions} summarize the findings, and provide an outlook of further applications and generalizations of the approach.

\section{Motivating example}
\label{sec:ssh}
\begin{figure}[t]
\includegraphics[width=\linewidth]{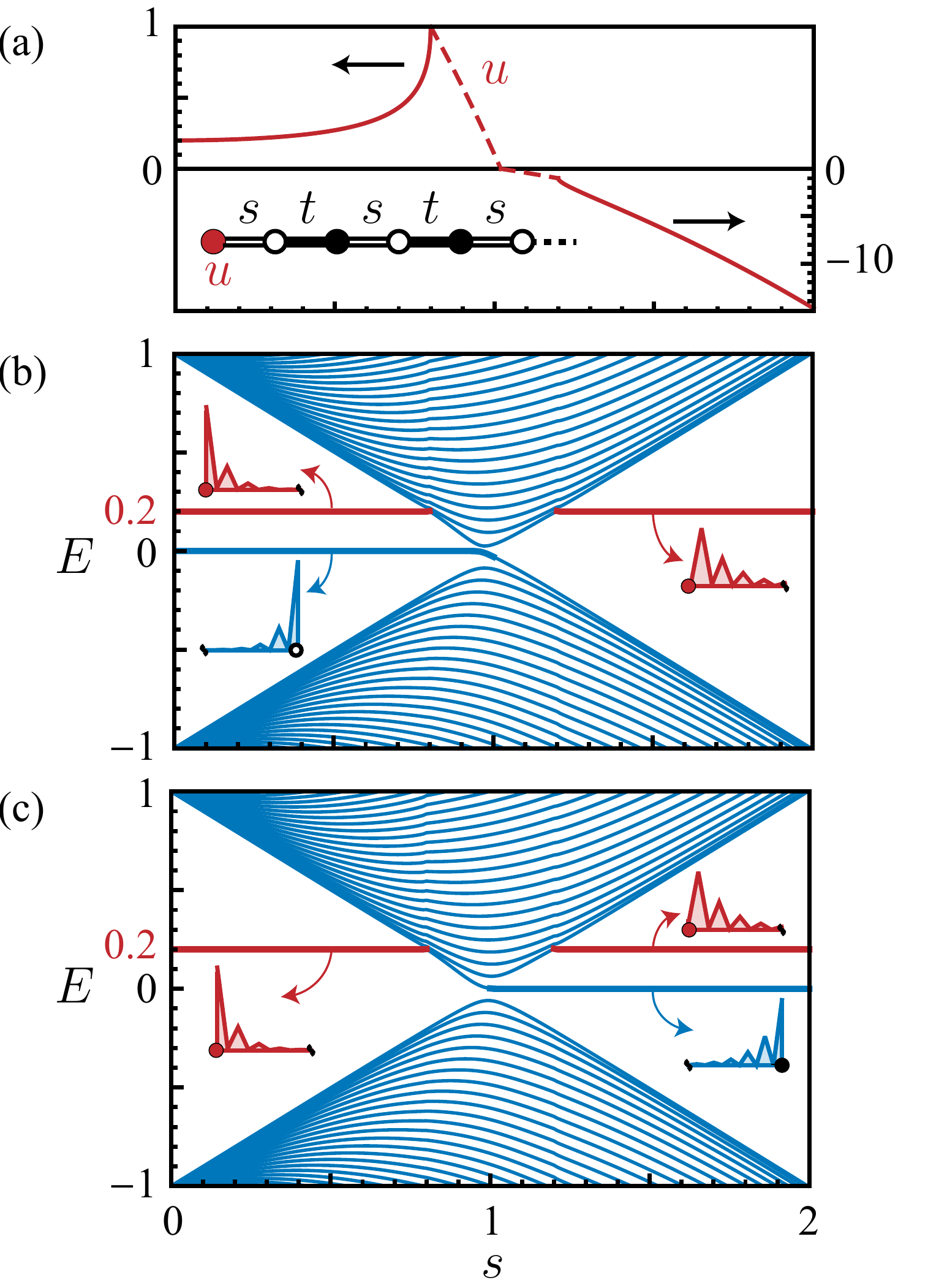}
\caption{Design of an interface state with a predetermined energy $E$ in a Su-Schrieffer-Heeger dimer chain with alternating couplings $s$ and $t$, irrespective of the topological phase of the bulk. (a) Required symmetry-breaking edge potential $u$ to induce a state of energy $E=0.2$, as a function of the coupling $s$ at $t=1$  (dashed: $\mathrm{Re}\,u$ while $E$ is outside the gap). The inset illustrates the system.
(b,c) Energy spectra of chains with $100$ and $101$ sites, with the induced interface state highlighted in red. Insets show intensity profiles on the 10 sites nearest to the edge at $s=0.5$ and $s=1.5$ (an extra zero mode localized at the other edge is also shown).}
\label{fig:ssh}
\end{figure}

As mentioned in the introduction, we prepare our general considerations by a motivating examples, where we equip a Su-Schrieffer Heeger chain with an analytically determined symmetry-breaking edge potential that induces a state of a predetermined, non-vanishing energy $E$.
In its original form \cite{Su79}, the Su-Schrieffer-Heeger chain is a Hermitian dimer chain with alternating couplings $s$ and $t$ and vanishing onsite potentials, which
can support edge states that are protected by a chiral symmetry. These states sit in the centre $E=0$ of the band gap $|E|<||s|-|t||$, and only appear in the topologically nontrivial phase, where $|s|<|t|$.
As shown in Fig.~\ref{fig:ssh}, we aim to design an edge unit cell with a symmetry-breaking potential $u$ that induces an edge state with a different, pre-determined, energy within the band-gap, including in the trivial phase.
This can be achieved by setting the edge potential to the nontrivial fixed point of the simple renormalization map
\begin{equation}
u'= t^2\frac{u - E}{s^2 + (u - E) E},
\label{eq:umap}
\end{equation}
which we can obtain from direct algebraic manipulations of the microscopic tight-binding model.
We derive this here in a form that facilitates its generalization to more complicated interfaces in the next section.

In the bulk, we write the tight-binding equations compactly as
\begin{equation}
E\psi_n=W\psi_{n-1}+H\psi_n+V\psi_{n+1},
\label{eq:sshbulk}
\end{equation}
where $\psi_n$ are two-component vectors containing the amplitudes on the $n$th dimer ($n=1,2,3,\ldots$), and
\begin{align}
H=\begin{pmatrix}
    0  & s \\
    s & 0\\
  \end{pmatrix},
  \qquad
V=W^T=\begin{pmatrix}
    0  & 0 \\
    t & 0\\
  \end{pmatrix}.
\end{align}
The chiral symmetry corresponds to the relations $\sigma_zA\sigma_z=-A$ for each of the matrices $A=H,V,W$, where $\sigma_z$ is a Pauli matrix.
At the edge, we allow for a symmetry-breaking onsite potential $u$,  corresponding to an internal edge Hamiltonian
\begin{equation}
H^{(0)}=\begin{pmatrix}
    u  & s \\
    s & 0\\
  \end{pmatrix},
\end{equation}
which enters the tight-binding equations coupling the edge dimer to the first bulk unit cell,
\begin{equation}
E\psi_0=H^{(0)}\psi_0+V\psi_1.
\end{equation}
We solve for
\begin{align}
\psi_0=(E-H^{(0)})^{-1}V\psi_1
\end{align}
and insert this into the equation \eqref{eq:sshbulk} for the adjacent unit cell ($n=1$) to obtain the equation
\begin{align}
E\psi_1=H^{(1)}\psi_1+V\psi_2,
\end{align}
where the renormalized effective edge Hamiltonian takes the form
\begin{equation}
H^{(1)}=H+W(E-H^{(0)})^{-1}V.
\end{equation}
This evaluates to
\begin{equation}
H^{(1)}=
\begin{pmatrix}
    u'  & s \\
    s & 0\\
  \end{pmatrix},
\end{equation}
where the renormalized edge potential
$u'$ is related to the bare edge potential by the map \eqref{eq:umap}.
This procedure can be iterated to generate a succession of renormalized edge Hamiltonians
\begin{equation}
H^{(n)}=H+W(E-H^{(n-1)})^{-1}V,
\end{equation}
with a corresponding succession of renormalized edge potentials.

\begin{figure}[t]
\includegraphics[width=\linewidth]{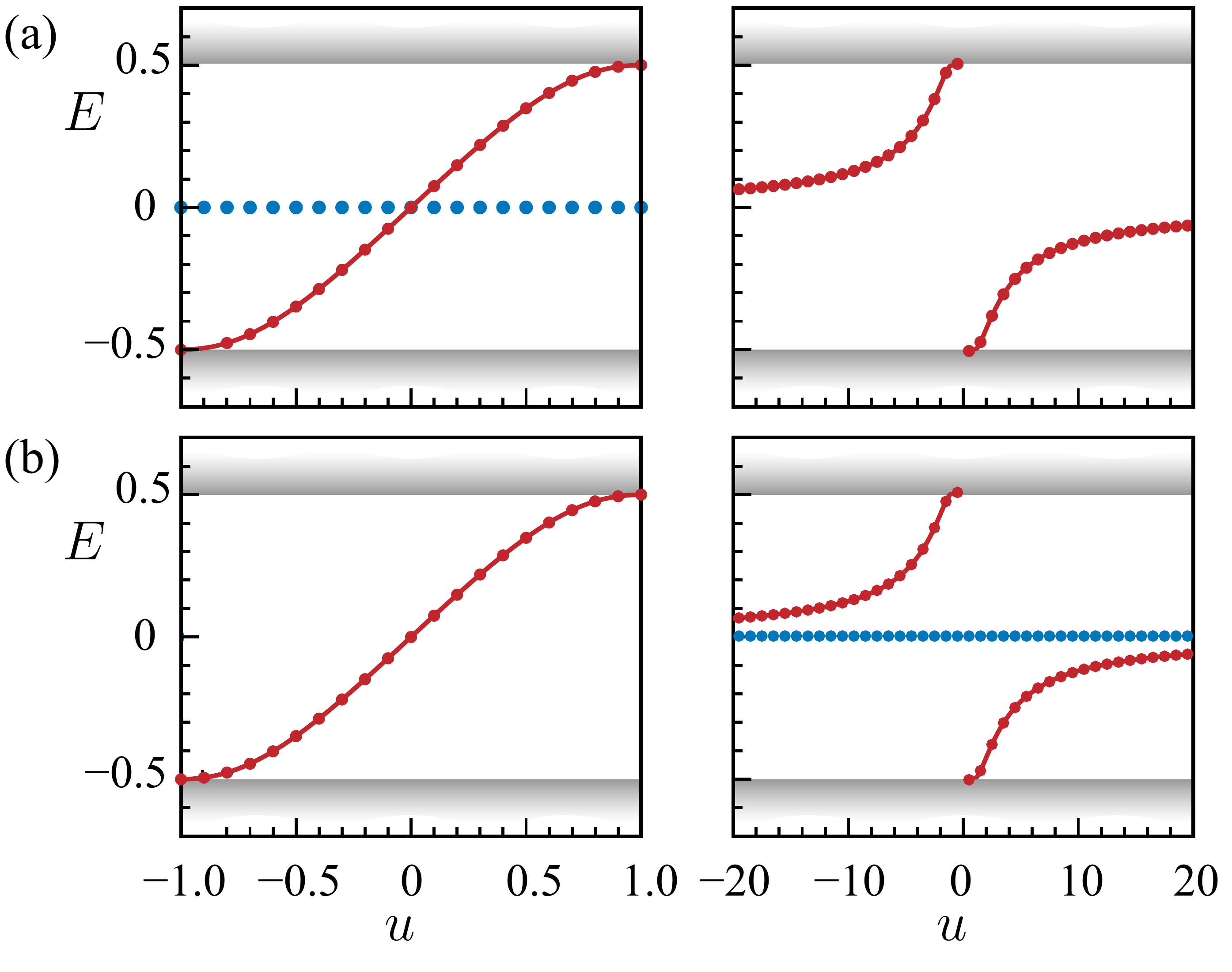}
\caption{Dependence of the energy $E$ of the induced interface state as a function of the symmetry-breaking potential $u$, in the SSH chain of Fig.~\ref{fig:ssh}. The curves represent the predictions obtained  from the unstable fixed point $u_-(E)$, Eq.~\eqref{eq:ufix}, of the renormalization map \eqref{eq:umap}, evaluated for
$s=0.5$, $t=1$ (left panels) and  $s=1.5$, $t=1$ (right panels).
The red data points are the numerical energies of the induced states, obtained from systems with (a) 100 sites and (b) 101 sites.
The blue dots in the top left and bottom right panels show the position of an additional near-zero mode localized at the unmodified far end of the system.}
\label{fig:ssh2}
\end{figure}

We note that the map \eqref{eq:umap} acts only on the edge potential, but its nature is fixed by the bulk parameters $s$ and $t$ as well as the energy $E$. In particular, this applies to the fixed points
\begin{equation}
u_\pm(E)=
\frac{E^2  +t^2 - s^2 \pm\sqrt{(E^2 -(t-s)^2)(E^2 -(t+s)^2)}}{2 E},
\label{eq:ufix}
\end{equation}
which are real for energies inside the gap.
From the stability coefficient
\begin{equation}
\left.\frac{du'}{du}\right|_{u_\pm}=\frac{s^2t^2}{[E(E-u^\pm)-s^2]^2} \equiv \lambda,
\label{eq:lambdassh}
\end{equation}
we find that within the gap $u_+$ is a stable fixed point ($|\lambda|<1$) and $u_-$ is an unstable fixed point ($|\lambda|>1$), so that the convergence to the latter is only obtain for specific energies $E$ or initial conditions $u$.

Critically, we find that this behaviour is directly linked to the quantization condition of the edge states.
We infer this from the asymptotic intensity profile
$||\psi_n||^2\propto |\mu|^{-2n}$ of the solutions, which we obtain from the
relation
\begin{equation}
\psi_{n}=(E-H^{(\pm)})^{-1}V\psi_{n+1}\equiv \mu \psi_{n+1}
\label{eq:musshdef}
\end{equation}
evaluated with the corresponding fixed-point Hamiltonian $H^{(\pm)}$.
The explicit calculation gives us
\begin{equation}
\mu=\frac{2st}{E^2  - s^2-t^2 \mp\sqrt{(E^2 -(t-s)^2)(E^2 -(t+s)^2)}},
\label{eq:mussh}
\end{equation}
which obeys the exact identity $\lambda=\mu^2$.
As we show in the next section, this identity is deeply imbedded into the foundations of the presented approach.
We conclude that a given fixed point supports a state with an asymptotic behaviour $||\psi_n||^2\propto |\lambda|^{-n}$ that is directly related to its stability coefficient, and is normalizable only if $|\lambda|>1$, hence if the fixed point is unstable.

In addition, we observe that within the bands, the fixed points $u_\pm$ are both complex, and hence cannot be approached from real initial conditions, so that the map does not converge to any fixed point at all. This is consistent with the behaviour of extended states that display a quasiperiodic beating pattern deep in the bulk. The band structure itself is recovered by writing $\mu=\exp(-ik)$ in terms of a dimensionless wave number $k$, with the sign following from the propagation direction implied by  Eq.~\eqref{eq:musshdef}.  Then Eq.~\eqref{eq:mussh} provides us with the known dispersion relation of the SSH model.

Therefore, we can obtain all information about the interface physics from the renormalization map.
For a fixed initial $u$, and a general energy $E$ inside the gap, the map converges to the stable fixed point $u_+$, which does not support a normalizable edge state.
The nontrivial edge physics is contained in the unstable fixed point $u_-$, to which one only converges for energies where an edge state exists.
In particular, we can solve the relation $u=u_-(E)$ for
\begin{equation}
E=\frac{t^2+u^2-\sqrt{4s^2u^2+(u^2-t^2)^2}}{2u}
\end{equation}
to obtain the energy of an edge state induced by a given edge potential $u$, as confirmed in Fig.~\ref{fig:ssh2}.

Furthermore, to achieve the goal set out at the beginning of this section, we can select  arbitrary values of $s$ and $t$, as well as a target energy $E$ within the range of the gap, and then determine a suitable edge potential $u=u_-(E)$ to induce an edge state at that energy. For $|s|<|t|$, this edge potential itself takes values inside the gap, while for $|s|>|t|$ it takes values outside the gap.  This edge potential is illustrated for $E=0.2$ in Fig.~\ref{fig:ssh}(a) as a function of $s$ for fixed $t=1$.
In a sufficiently long system, this then results in an edge state with the desired energy, which can be achieved irrespective of the topological phase of the system. This is verified in Figs.~\ref{fig:ssh}(b,c), where we show how the energy spectrum of finite systems of 100 and 101 sites develops, again for fixed $t=1$ and variable $s$.
Depending on the phase, this state can then be interpreted as a deformation of an edge state that preexisted before introducing the edge potential ($|s|<|t|$), or a state drawn out of the bulk where such a predecessor did not exist ($|s|>|t|$).
As also illustrated in this figure, this state can coexist with a second edge state localized at the opposite end of the system, which has an energy very close to the band centre and occurs only in the phase where the far edge is in the nontrivial configuration ($|s|<|t|$ if the system size is even and $|s|>|t|$ if the system size is odd).

\section{General approach}
\label{sec:general}
\subsection{Strategy}
To develop the interface renormalization approach in its general form, we place the methodology into the context of a simple and concrete algorithmic framework, involving the recursive calculation of Greens functions in terms of a Dyson series \cite{datta1997electronic}.
Via this approach, parts of the system can be successively eliminated, which is achieved by relating them algebraically to their coupled neighbors, and is ideally suited to describe settings that afford a quasi-one-dimensional periodic stratification.
Here, we seek to design a specific iterative procedure that leads to an effective theory of interface states, and aim to exploit its
mathematical structure as a map in the interface parameter space.
As illustrated in Fig.~\ref{fig:generalapproach},
this can be achieved by systematically absorbing a unit cell of the bulk into the interface, so that the bulk returns into its original configuration while the interface parameters are renormalized. The natural general setting are therefore systems with periodic bulk regions joined up by interfaces, as well as the edges of such systems.
Indeed, for general considerations it suffices to study the edge configuration, sketched in Fig.~\ref{fig:generalapproach}(a), as interfaces between regions can  be mapped to it as shown in Fig.~\ref{fig:generalapproach}(b). The result will be a map in an effective interface parameter space, as shown in Fig.~\ref{fig:generalapproach}(c), which we can analyze using dynamical systems theory.

\subsection{Implementation}
This strategy can be put into the following concrete mathematical form. We consider an edge region (index $n=0$) with an internal Hilbert space of dimension $N_0$,
coupled to the bulk by possibly nonreciprocal and effectively non-Hermitian matrices $v$ and $w$, and equipped with a potentially non-Hermitian effective internal Hamiltonian $h$.  In the bulk, we adopt a periodic arrangement of super-cells (index $n=1,2,3,\ldots$), with an internal dimension $N$ that is chosen large enough so that all bulk and edge couplings will be to neighboring cells only. The super-cell may therefore contain several fundamental unit cells. The bulk is then completely described by a internal Hamiltonian $H$  and couplings $V$ and $W$ to the neighboring super-cells, which can again be non-Hermitian and nonreciprocal \footnote{In higher-dimensional models with periodicity along the interface, all these matrices may be replaced by Bloch versions that are parameterized by transverse wavenumbers or vectors.}.
This setup translates to the coupled-mode equations
\begin{equation}
E\mathbf{\psi}_0=h\mathbf{\psi}_0+v \mathbf{\psi}_1
\end{equation}
on the edge cell,
\begin{equation}
E\mathbf{\psi}_1= w\mathbf{\psi}_0+H\mathbf{\psi}_1+ V\mathbf{\psi}_2
\end{equation}
on the first bulk supercell, and
\begin{equation}
E\mathbf{\psi}_n= W\mathbf{\psi}_{n-1}+H\mathbf{\psi}_n+ V\mathbf{\psi}_{n+1}
\end{equation}
in the remainder of the bulk.

In a first step, we solve for
\begin{equation}
\mathbf{\psi}_0=(E-h)^{-1}v\mathbf{\psi}_1
\end{equation}
and insert this into the first bulk equation to write
\begin{align}
E\mathbf{\psi}_1&= H^{(1)}\mathbf{\psi}_1+ V\mathbf{\psi}_2,
\end{align}
with renormalized Hamiltonian
\begin{align}
H^{(1)}&=[H+w(E-h)^{-1}v] \equiv m(h)
\label{eq:preproc}
\end{align}
on the new effective edge.
We view this as a preparatory
step as it changes the internal edge-space dimension from $N_0$ to $N$. From here on, we can iterate the procedure with a fixed internal dimension $N$, where the leading supercell plays the role of the renormalized edge.
This iteration takes the form
\begin{align}
E\mathbf{\psi}_n&= H^{(n)}\mathbf{\psi}_n+ V\mathbf{\psi}_{n+1},
\end{align}
where
\begin{align}
H^{(n)}&=[H+W(E-H^{(n-1)})^{-1}V]\equiv M(H^{(n-1)})
\label{eq:genmap}
\end{align}
is the desired effective renormalization map in the interface parameter space.

\subsection{Structural aspects and simplifications}

We note that the renormalization map \eqref{eq:genmap} resembles a composition rule of self-energies, in analogy to the Dyson series in recursive Green's functions employed in transport calculations \cite{datta1997electronic}. The key aspect is the nonlinearity of this rule. In transport calculations, this provides intrinsic numerical stability that is not observed, e.g., in transfer-matrix calculations.
In the present context, where we deal with initial conditions that describe interfaces, this nonlinearity allows us to characterize the resulting states in the language of nonlinear systems, and use such insights also for the design of such states. Before we establish this in detail, we point out some other useful structural features of the renormalization map.

For this, we note that this map only depends on the bulk parameters and the energy, and hence will reflect any further bulk symmetries, even when these are broken by the interface that provides the initial conditions.
For instance, the short-range couplings encountered in the most ubiquitous models yield matrices $V$ and $W$ of a low rank, so that the number of renormalized interface parameters is automatically small. Short range couplings also allow to efficiently employ block inversion formulas, such as
\begin{equation}
\begin{pmatrix}
    A  & B \\
    C & D\\
  \end{pmatrix}^{-1}_{\mathrm{top~left~block}}=(A+BD^{-1}C)^{-1},
\end{equation}
which can be reinterpreted as decomposing the map into smaller steps.
If the matrices are of a more general block structure, such as in the mapping of the interface on an edge in Fig.~\ref{fig:generalapproach}(b), the renormalization of interface parameters will still be accordingly restricted. Finally, symmetries of topological significance, such as chiral or charge conjugation symmetries, further constrain the renormalization flow, and lead to dualities in parameter space that connect equivalent interfaces, even when these break the symmetries---we list these constraints and  dualities for particularly common cases in Appendix \ref{app:sym}.
Because of these ubiquitous features, we typically obtain a simple and effective interface theory, which can be analysed in specific models as illustrated further below.
In all cases, the interpretation of the theory rests on a direct conceptual duality between its dynamical-system aspects and the microscopic quantization condition of the interface states, which we establish next.

\begin{figure*}[t]
\includegraphics[width=\linewidth]{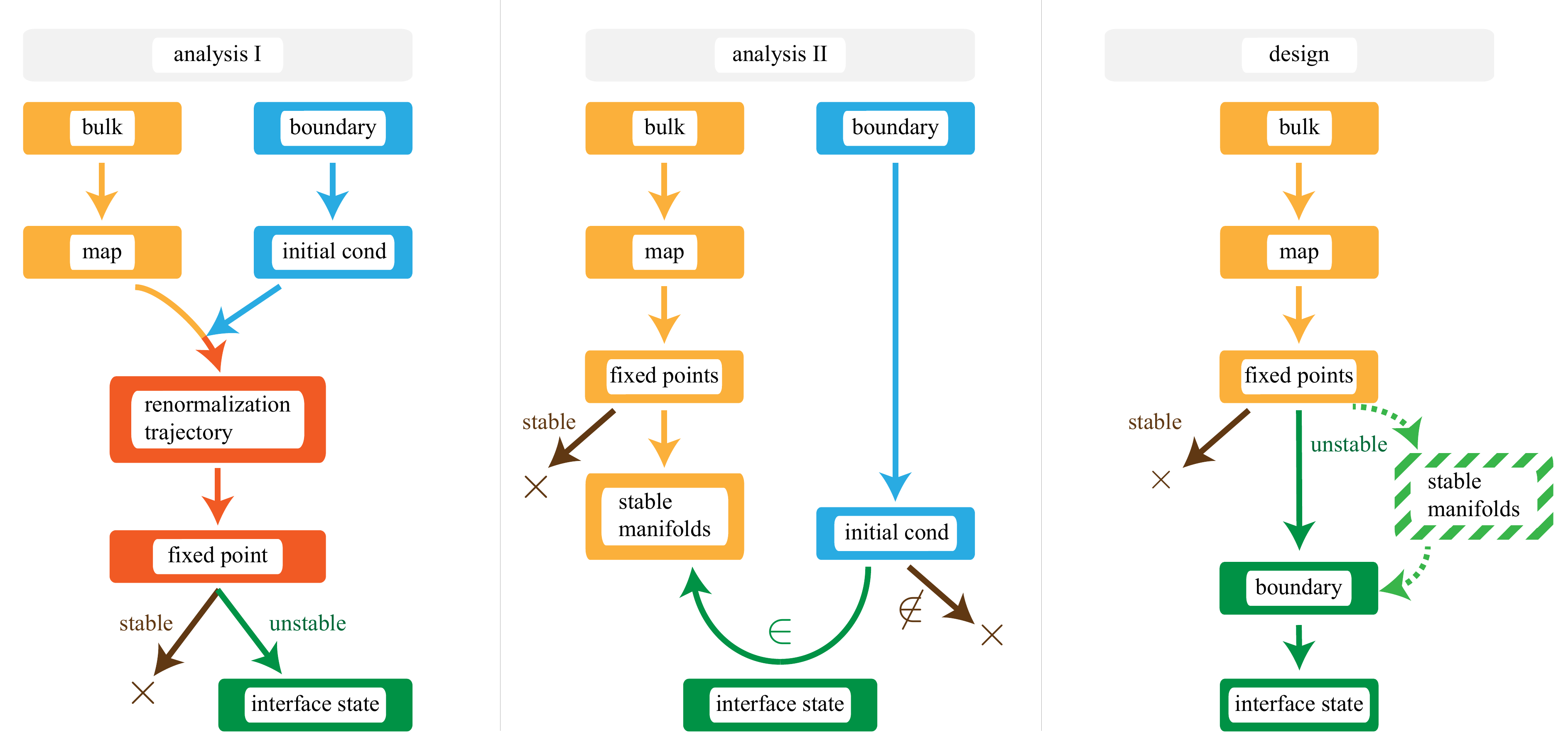}
\caption{Interface analysis and design principles. In all cases, the bulk data determines the renormalization map, its fixed points, and their unstable manifolds, whose complexity is often simplified by the bulk symmetries.
Practically, the existence of an interface state at a given energy and interface configuration can then be asserted by inspecting whether the appropriately initialized renormalization trajectory converges to an unstable fixed point (analysis I). This is equivalent to determining whether the initial condition lies on the stable manifold of such a fixed point, resulting in a clear separation of the bulk and boundary data (analysis II). For the design of interfaces, the required interface configuration can be obtained directly from an unstable fixed point, or any point along its stable manifold (design).
}
\label{fig:generalapproach2}
\end{figure*}

\subsection{Stability versus quantization}
\label{sec:stabquant}
We now describe an important technical feature underpinning the presented approach. This concerns the direct link between the stability of the map and the quantization condition of the interface states. Based on this link, we will be able to determine the existence of interface states by studying the renormalization map alone, without needing to refer back to the original wave equation.

We establish this link by inspecting the asymptotic behaviour of the states deep in the bulk, hence, after many renormalization steps,
which we assume to be governed by a fixed point $H^{(\infty)}$ of the map.
In the general setting of potentially non-reciprocal non-Hermitian systems,
this generates \emph{two} recursion relations,
\begin{align}
\psi_n&=(E-H^{(\infty)})^{-1}V\psi_{n+1},\nonumber\\
\varphi_n&=\varphi_{n+1}W(E-H^{(\infty)})^{-1}
\label{eq:recrel}
\end{align}
one for the left wavefunctions $\psi_n$ and one for the right wavefunctions $\varphi_n$.
The latter obey the wave equation
\begin{equation}
E\varphi_n= \varphi_{n-1}V+\varphi_n H+\varphi_{n+1}W,
\end{equation}
from which we obtain the reduced equations
\begin{equation}
E\varphi_n=\varphi_n H^{(n)}+\varphi_{n+1}W
\end{equation}
with exactly the same sequence \eqref{eq:genmap} of renormalized Hamiltonians as for the right wavefunctions $\psi_n$.

With the recursion relations \eqref{eq:recrel}, the asymptotic solutions can be decomposed into states obeying
\begin{align}
\label{eq:eqdec1}
\mu^{(\alpha)}\psi^{(\alpha)}&=(E-H^{(\infty)})^{-1}V\psi^{(\alpha)},\\
\nu^{(\beta)}\varphi^{(\beta)}&=\varphi^{(\beta)}W(E-H^{(\infty)})^{-1},
\label{eq:eqdec2}
\end{align}
which induce an exponential spatial profile governed by the factors $\mu^{(\alpha)}$ and $\nu^{(\beta)}$,
\begin{align}
\psi_n&\propto \psi^{(\alpha)}{\mu^{(\alpha)}}^{-n},\\
\varphi_n&\propto \varphi^{(\beta)}{\nu^{(\beta)}}^{-n}.
\end{align}
According to the general biorthogonal quantization theory in nonreciprocal systems \cite{Kun18}, a normalizable interface state then arises when we can construct from this a biorthogonal pair of specific solutions, which we give the superscript $(0)$, so that
\begin{equation}
|\mu^{(0)}\nu^{(0)}|>1.
\label{eq:biorthquantcond}
\end{equation}

Our key observation is that this intricate mathematical condition is directly reflected in the stability properties of the map \eqref{eq:genmap}.
Let us write it over one iteration as $M(\mathcal{H})=\mathcal{H}'$. The Jacobian matrix
\begin{equation}
J_{kl,mn}=\frac{\partial \mathcal{H}'_{kl}}{\partial \mathcal{H}_{mn}}=(W(E-\mathcal{H})^{-1})_{km}((E-\mathcal{H})^{-1}V)_{nl}
\end{equation}
then factorizes across the whole parameter space, and exactly into the two operators that determine the existence of interface states. Evaluating this at a fixed point $H^{(\infty)}$, we can construct eigenstates as
\begin{equation}
\mathcal{V}_{kl}^{(\alpha\beta)}=\varphi^{(\beta)}_k \psi^{(\alpha)}_l.
\label{eq:linmapevecs}
\end{equation}
According to
\begin{equation}
\sum_{kl}\mathcal{V}_{kl}^{(\alpha\beta)}J_{kl,mn}=\mu^{(\alpha)}\nu^{(\beta)}\mathcal{V}_{mn}^{(\alpha\beta)},
\end{equation}
the corresponding stability eigenvalues are given by $\mu^{(\alpha)}\nu^{(\beta)}$.
Therefore, the linear stability of the fixed points in the renormalization map gives us direct access to the quantization condition \eqref{eq:biorthquantcond} of the full wave equation.

In practice, this means that normalizable interface states can be found for convergence to nontrivial fixed points with unstable directions. The generic nontrivial fixed points will have one unstable direction only, so that they can be located by changing only one parameter, leading to a discrete quantization of their energies, while fixed points with more unstable directions describe degeneracies, which can signal phase transitions.
Notably, this link applies  irrespective of whether the system is Hermitian or non-Hermitian, and whether it obeys a traditional bulk-boundary correspondence or not. Applied to concrete models, this becomes a very efficient procedure not only for the analysis but also for the design of interfaces, as we describe next.

\subsection{Design of interface states}
\label{sec:design}
The interface quantization procedure of the previous subsection can be reinterpreted to design interfaces that support states with desired properties.
In particular, fixing the bulk parameters and a desired interface-state energy fixes the general map, and hence also the nontrivial fixed points.
An  interface state with the desired energy is then obtained for any parameter combination on the stable manifold of these fixed points, or more directly for just the fixed-point parameters themselves.
The preimages of these structures under the simple algebraic map $m$ translate this data into the space of bare interface parameters $h$.

As an important corollary, this construction can be made exact in finite systems provided we also allow for free control of the far end of the system. Setting $h=H^{(\infty)}$ to an unstable fixed point at the specified energy, exact solutions of the wave equation  are obtained when the far end is equipped with an interface Hamiltonian $h'=H+E-h$.
This then automatically fulfills the fixed point condition also for iteration from the second interface at the far end.
The result is a maximal number of exact degenerate interface states with precisely the prescribed energy $E$ in a finite system. Furthermore, these solutions display a purely exponential envelope decaying away from one of the interfaces, and hence realize the asymptotic considerations of Sec.~\ref{sec:stabquant} in an exact form.

\subsection{Summary of renormalization principles}
\label{sec:principles}
Before we turn to illustrate the renormalization analysis and design of interfaces in concrete examples we summarize the key general characteristics of the obtained effective theory, whose interdependencies are further illustrated in Fig.~\ref{fig:generalapproach2}.
\begin{itemize}
\item[]\emph{Bulk-boundary separation.}
The map \eqref{eq:genmap} itself only depends on the bulk parameters and energy, but its initial conditions (potentially preprocessed as in Eq.~\eqref{eq:preproc}) are determined by the bare interface parameters, which are renormalized in each iteration of the map.
\item[]\emph{Parameter compression.}
Bulk symmetries further reduce the number of relevant parameters, both through relations between them as well as through conservation laws. This includes symmetries of topological significance,
and results in systematic simplifications in the effective interface theory.
A similar reduction also occurs when bulk couplings are of short range or structured into blocks, which reduces the number of renormalized parameters and  allows to decompose the map into simpler steps.
As we will see in the examples, the effective interface dimension of paradigmatic models is typically very small.
\item[]\emph{Universal description of interface states.}
Edge and interface states are characterized by convergence of their parameters to a non-trivial fixed point with unstable directions.
For fixed system parameters, this becomes a search for energies where the bare parameters lie on the stable manifold of such a fixed point. The analysis of interface states is therefore tied to dynamical-systems theory in a low-dimensional space. This link applies irrespective of whether the system is Hermitian or non-Hermitian, and whether it obeys a traditional bulk-boundary correspondence or not.
\item[]\emph{Universal design principle of interface states.}
The universal quantization principle can be reinterpreted to design interfaces that support states with desired properties, and in particular states with a predetermined energy, for which one can use the parameters of nontrivial unstable fixed points or their stable manifolds.
This interface construction principle directly addresses the design parameters while leaving  freedom to control other desired aspects of the system, such as the bulk band structure.

\end{itemize}

\begin{figure*}[t]
\includegraphics[width=\linewidth]{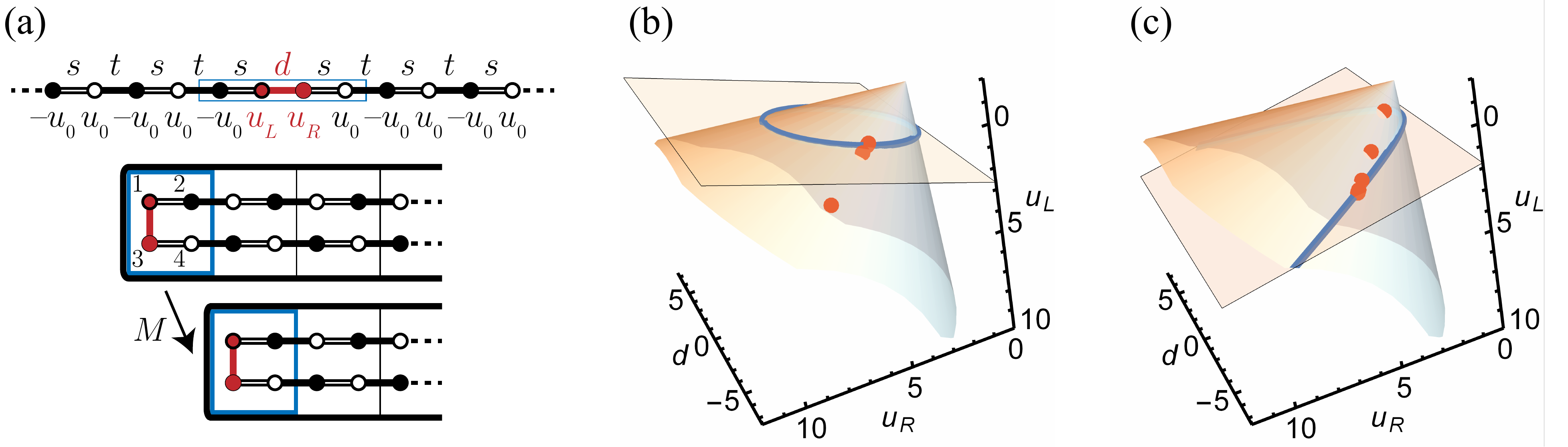}
\caption{(a) Renormalization structure in a generally non-Hermitian dimer chain (alternating couplings $s$ and $t$ and bulk potentials $\pm u_0$), linked by a defect dimer with bare coupling $d$ and onsite potentials $u_L$ and $u_R$. These parameters are the only ones that are renormalized by the map $M$, Eqs.~\eqref{eq:mapdimer}, \eqref{eq:fullmap}, leading to an effectively three-dimensional description of the interface. As shown in panels (b) and (c), the nontrivial fixed points form sections on their conical stable manifolds, which are anchored at a fixed point describing a disconnected system. In panel (b), where $s=1/2$, $t=1$, $u_0=0.1$, and $E=0.2$, the section is an ellipse, while in panel (c), where $u_0$ is changed to 0.3, the section is a hyperbola (the second branch of which is not shown). The dots represent a sample trajectory converging to such a nontrivial fixed point.}
\label{fig:dimer}
\end{figure*}

\subsection{Scope of illustrative applications}

Compared to the motivating example of Sec.~\ref{sec:ssh}, the general approach can be used to analyze and design much more complicated bulk and interface configurations.
For instance, the general principles can be systematically exploited to draw up phase diagrams of the interface states irrespective of any topologically relevant bulk symmetries, and efficiently design states with predetermined properties also in the complex energy plane. In the following sections, we demonstrate this for systems with broken bulk symmetries, non-Hermitian phase transitions, and non-reciprocal couplings, and cover cases where the interfaces separate two bulk regions or form the boundary of a higher-dimensional system.
The first application, a generalization of the motivating example, is designed to illuminate the general renormalization principles in a sufficiently complex but analytically tractable setting.
This focuses more on the detailed theoretical aspects of the approach, while subsequent examples focus more practically on the efficient design of physical interface phenomena.

\section{Non-Hermitian dimer chains linked by a defect}
\label{sec:dimer}

\subsection{Set-up and motivation}
To obtain a deeper analytical understanding of the general approach, we use it to extend the considerations of the motivating example to dimer chains with non-Hermitian, symmetry-breaking potentials, placed into an interface configuration with a general defect dimer as depicted in Fig.~\ref{fig:dimer}.
The two seminfinite bulk segments again feature staggered couplings $s$ and $t$, but now also include a staggered, possibly complex onsite potential $\pm u_0$  (that this averages to zero can always be achieved by a suitable choice of the reference energy).
In the Hermitian limit, where all parameters are real, this results in a gapped band structure. For finite $u_0$, the system  corresponds to a general Rice-Mele model \cite{Rice1982}, while for  $u_0=0$ it specializes to the Su-Schrieffer-Heeger model studied above.
For complex $u_0$, the imaginary parts of the onsite potentials correspond to gain and loss and introduce nontrivial non-Hermitian bulk effects \cite{Ram10,Lang18,Mostafavi2020}. In particular, when the gain-loss contrast is large enough, the band gap closes and complex band branches emerge. We are interested in the interplay of these intricate bulk features with states induced by the interface, which consists of a defect dimer with modified, possibly symmetry-breaking onsite potentials  $u_L$, $u_R$ and a distinct intra-dimer coupling $d$.
This more complicated set-up allows us to illustrate the joint renormalization of several interface parameters while incorporating a range of special cases that highlight the role of symmetry-breaking and non-Hermitian terms.

\subsection{Derivation of the renormalization map}
To implement the general approach,
we set the interface data to
\begin{align}
h=
\begin{pmatrix}
    u_L & s &  d & 0 \\
    s & -u_0  &  0 & 0\\
    d & 0 & u_R & s \\
    0 & 0 & s & u_0 \\
  \end{pmatrix},
  \quad v=w^T=
  \begin{pmatrix}
    0 & 0 & 0 & 0 \\
    t & 0 & 0 & 0 \\
    0 & 0 & 0 & 0 \\
    0 & 0 & t & 0
  \end{pmatrix}.
\end{align}
and the bulk data to
\begin{align}
H=\begin{pmatrix}
    u_0 & s &  0 & 0 \\
    s & -u_0  &  0 & 0\\
    0 & 0 & -u_0 & s \\
    0 & 0 & s & u_0 \\
  \end{pmatrix},
  \quad V=W^T=v.
\end{align}
We have chosen the interface region so that we can directly apply the
renormalization map \eqref{eq:genmap} with initial condition $ H^{(0)}\equiv h$, meaning that we work directly in the space of the interface parameters.
This generates a sequence of renormalized interface Hamiltonians
 \begin{equation}
H^{(n)}=\begin{pmatrix}
    u_L' & s &  d' & 0 \\
    s & -u_0  &  0 & 0\\
    d' & 0 & u_R' & s \\
    0 & 0 & s & u_0 \\
  \end{pmatrix},
  \label{eq:dimerh1}
\end{equation}
 of the same structure as $h$,
where the primed parameters are determine by consecutive application of a map
\begin{equation}
(d',u_L',u_R')=M(d,u_L,u_R)
\end{equation}
in the three-dimensional defect parameter space.
This map can be conveniently specified as a composition of two steps,
\begin{equation}
M= m_{-u_0,t}\circ m_{u_0,s}.
\label{eq:mapdimer}
\end{equation}
where the map
\begin{align}
m_{u,r}:\qquad
d'&=\frac{r^2d}{(E-u_L)(E-u_R)-d^2},
\nonumber
\\
u_L'&=-u+\frac{r^2(E-u_R)}{(E-u_L)(E-u_R)-d^2},
\nonumber
\\
u_R'&=u+\frac{r^2(E-u_L)}{(E-u_L)(E-u_R)-d^2}
\label{eq:fullmap}
\end{align}
acts in the same space,
and the subscripts keep track of the bulk parameters.

\subsection{Application of renormalization principles}
To characterize the interface via the map $M$,  we make use of the  principles listed in Sec.~\ref{sec:principles}.
We verify immediately that in accordance with the general bulk-boundary separation, the bulk parameters fix the nature of this map but do not undergo renormalization.
Furthermore, as we show in Appendix \ref{sec:appdimer}, the decomposition of the map into two steps arises naturally when we utilize the short-range nature of the bulk couplings in the renormalization procedure.
Let us next inspect how the renormalization procedure is simplified by the symmetries of the problem.

\subsubsection{The role of symmetries}

Any bulk symmetries of the Hamiltonian, including those of topological relevance, can be translated into mappings in the renormalization parameter space, so that these symmetries are reflected in the map, too. Let us establish for a specific example how this can be used to simplify the map.
In the present case, we observe that for $u_0=0$, where the bulk obeys a chiral symmetry,
\begin{align}
\frac{u_L'-u_R'}{2d'}=\frac{u_L-u_R}{2d}\equiv A
\label{eq:conserved}
\end{align}
defines a conserved quantity, which captures the amount of chiral symmetry breaking induced by the defect.
This conservation law allows us to reduce the single-step map $m_{u,r}$ to a map
\begin{align}
\bar m_{r}:\qquad d'&=\frac{r^2d}{(E-\bar u)^2-d^2(1+A^2)},
\nonumber\\
\bar u'&=\frac{r^2(E-\bar u)}{(E-\bar u)^2-d^2(1+A^2)}
\label{eq:dimermapreduced}
\end{align}
 in a two-dimensional parameter space,
where $\bar u=(u_L+u_R)/2$.
For the renormalization, we again compose this into a map
\begin{equation}
\bar M=\bar m_{t}\circ\bar m_{s}.
\label{eq:dimermapreduced2}
\end{equation}
Therefore, the chiral symmetry in the bulk reduces the complexity of the map significantly, in this case via a conservation law, and this can be exploited even if the interface breaks this symmetry.

\subsubsection{Relation of fixed points to interface states}

We now utilize the map $M$ to describe the interface states in the chain. For this, we seek the energies for which the map allows \emph{convergence to a nontrivial fixed point}.

Let us establish what the highlighted phrase means in the present context.
As shown in Fig.~\ref{fig:dimer}, the nontrivial fixed points of $M$ form elliptic or hyperbolic curves in the three-dimensional defect parameter space, while convergence occurs for initial conditions on the stable manifold of this curve.
We find that this stable manifold forms a conical surface, which is anchored at the fixed point
\begin{align}
X_-^{(0)}:\quad d=0,\quad u_L=\alpha_+-\beta_+\quad u_R=\alpha_--\beta_-,
  \label{eq:trivfp}
\end{align}
where
\begin{align}
\alpha_\pm&=\frac{(E\pm u_0)^2+t^2 -s^2}{2(E\pm u_0)},
\\
\beta_\pm&=
\frac{\sqrt{(E^2 - (s - t)^2 - u_0^2) (E^2 - (s + t)^2 - u_0^2)}}{2(E\pm u_0)}.
  \end{align}
The anchoring fixed point itself  describes the decoupled system, and is unstable in all directions, so that is not accessible from the initial interface configuration.
Generically, the control of a single parameter interface parameter is then required to place an initial condition onto the conical stable manifold of the nontrivial fixed points on the ellipse or hyperbola.

To illuminate this structure further, we specialize to the case $u_0=0$ described by the map  $\bar M$.
We find the following four fixed points of $\bar M$,
\begin{align}
&X_\pm^{(0)}:\quad d_\pm^{(0)}=0,\quad\bar u_\pm^{(0)}=\alpha\pm \beta,
\\
&X_\pm^{(1)}:\quad d_\pm^{(1)}=\pm\frac{\beta}{\sqrt{1+A^2}},\quad\bar u_\pm^{(1)}=\alpha,
\end{align}
where
\begin{align}
\alpha=\frac{E^2+t^2-s^2}{2E},\quad
\beta=\frac{\sqrt{(E^2 -(t-s)^2)(E^2 -(t+s)^2)}}{2E}
\label{eq:ab}.
\end{align}
The two fixed points  $X_\pm^{(0)}$ again describe situations in which the defect coupling flows to zero, and the effective defect potential is independent of the bare defect potentials, erasing all signatures of its existence. The system then effectively reduces to two copies of the motivating example. This is reflected in the values of the $u_\pm^{(0)}$, which agree with Eq.~$\eqref{eq:ufix}$.
We already established that the fixed point  $X_+^{(0)}$ is trivial. The fixed point $X_-^{(0)}$ coincides with the one specified in Eq.~\eqref{eq:trivfp}. Its the stable manifold is just the point itself, so that it is only accessible when the system is decoupled from the very beginning.

In contrast, the two fixed points $X_\pm^{(1)}$ describe  collective modes with a finite effective coupling  $d_\pm^{(1)}$ that furthermore depends on the bare defect potentials. They only differ in the sign of this coupling, which means that they can be mapped onto each other by inverting the wave amplitudes in one half of the system.
These are therefore  nontrivial fixed points that describe genuine interface states, of even or odd parity.

In order to converge to a given nontrivial fixed point, the initial conditions of $d$ and $\bar u$, determined by the bare defect parameters, must again lie on its stable manifold. In the present special case, this stable manifold forms a straight line connecting it to the decoupling fixed point $X_-^{(0)}$. We parameterize this line by a scalar $\delta$,
\begin{equation}
(d(\delta),\bar u(\delta))=(0, \alpha-\beta)+\delta (1, \pm\sqrt{1+A^2}),
\label{eq:stabman}
\end{equation}
so that it runs through the nontrivial fixed point at $\delta =d_\pm^{(1)}$, while the decoupling fixed point $X_-^{(0)}$ is passed at $\delta=0$.

To identify the energy that places the bare-coupling initial conditions  onto this line, we need to simultaneously fulfill the conditions
$d(\delta)=d$, $\bar u(\delta)=\bar u$. This fixes $\delta=d$ and furthermore requires
\begin{equation}
u_L+u_R \mp  \sqrt{4 d^2+(u_L-u_R)^2}=2(\alpha-\beta).
\end{equation}
In this condition, the left-hand side only involves the bare defect parameters, which we have written out explicitly, while the right-hand side only involves bulk parameters and the energy, see Eq.~\eqref{eq:ab}.
This then determines the energies of the interface states, in this case as an explicit, if lengthy, algebraic expression. For conciseness, we give this here for the case $u_L=-u_R\equiv u$, where the interface-state energies take the form
\begin{align}
E_\pm=\pm\frac{t^2+d^2+u^2-\sqrt{4s^2(d^2+u^2)+(t^2-d^2-u^2)^2}}{2\sqrt{d^2+u^2}}.
\label{eq:esym}
\end{align}

\begin{figure}[t]
\includegraphics[width=\linewidth]{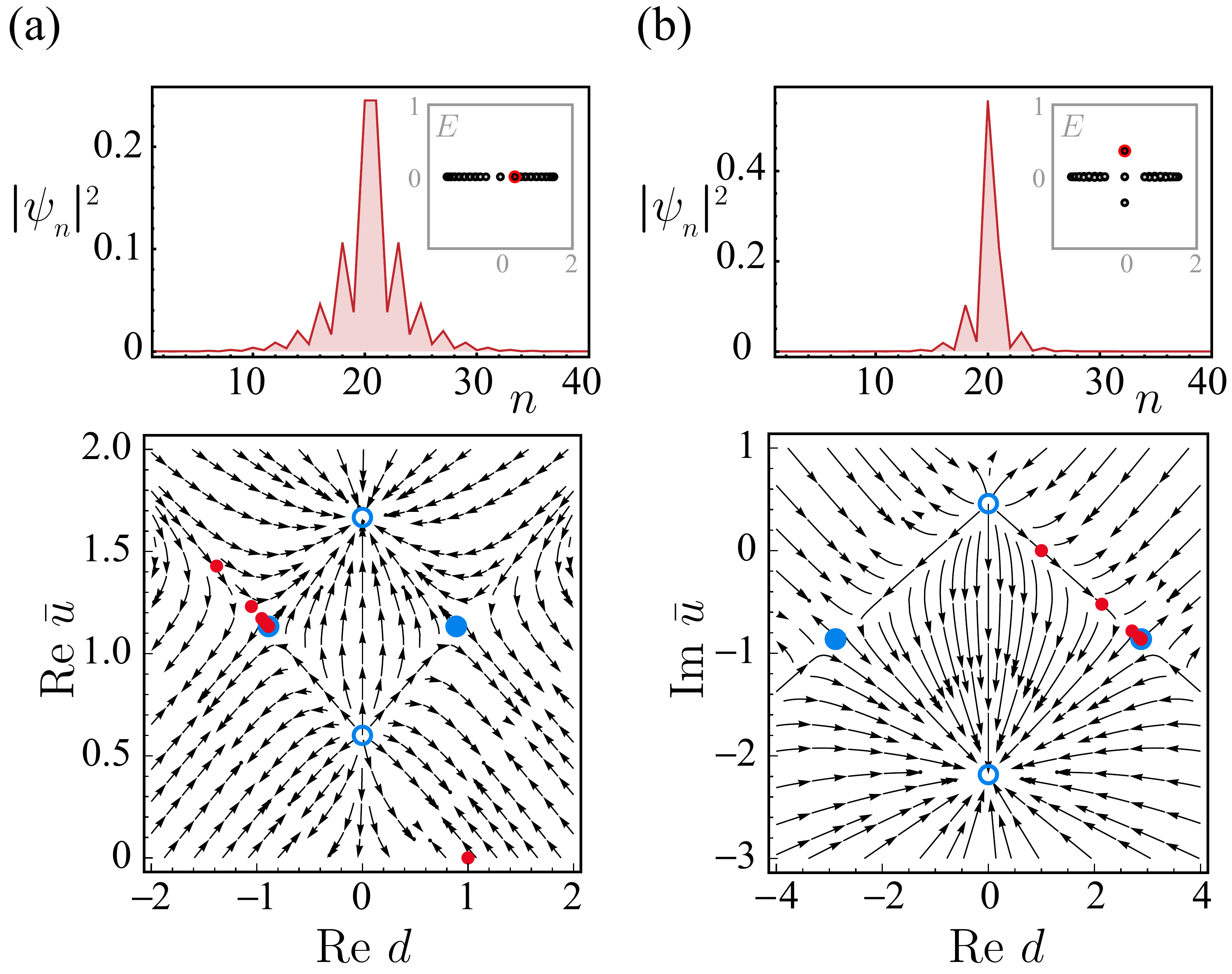}
\caption{Description of interface states in the dimer chain of Fig.~\ref{fig:dimer} with vanishing bulk potential $u_0=0$, as governed by the reduced renormalization map $\bar M$, Eqs.~\eqref{eq:dimermapreduced}  and \eqref{eq:dimermapreduced2}. We set $s=1/2$, $t=d=1$ and contrast the cases
(a) $u_L=-u_R=0.8i$ (hence $A=0.8i$) and (b) $u_L=-u_R=1.1i$ (hence $A=1.1i$), which place the system into the PT exact and the PT broken phase.
The bottom panels show the relevant cross-sections of the corresponding renormalization flow (arrows)
in the complex interface parameter space, where the energy is set to the predicted value (a) $E_+\approx 0.4020$ and (b) $E_+\approx 0.3599 i$ of an interface state, Eq.~\eqref{eq:esym} with $u=A$ (the flow at the energy $E=E_-$ of the second interface state is obtained by sending $\bar u\to -\bar u$).
This confirms that  starting from the initial condition $(d,\bar u)=(1,0)$ set by the bare interface parameters, the renormalization trajectory (red dots) then converges to a nontrivial fixed point (full blue dots), while a trivial fixed point (open dots) is approached at general energies or initial conditions.
The top panels show the intensity profiles $|\psi_n|^2$ of the associated interface states in a chain of 20 dimers (40 sites indexed by $n$), with the inset indicating their position in the complex energy spectrum.
This moderate system size is sufficient so that the interface states are well localized within the system, despite them being energetically close to the extended states (a)  or offset by a similar amount into the complex plane (b). Furthermore, the interface only minimally affects edge states at the far end of the system, whose energy is close to 0.}
\label{fig:flow}
\end{figure}

\subsubsection{Complex phase diagrams}

A useful feature of the described approach is that it directly unfolds in the interface
parameter space, and holds across all of it, including for complex parameters. We therefore can utilize this approach
to further divide the space into phases where the interface states take on different characteristics, including due to non-Hermitian effects. In the present example, the phase boundaries are determined by branch cuts that signal reconfigurations of the fixed points in
parameter space. For instance, for real parameters, the fixed point manifolds change from elliptic to hyperbolic at $E=u_0$, as already illustrated in Fig.~\ref{fig:dimer}.

The same principles also hold for complex parameters, but then lead to a much more complicated phase diagram.
Consider, e.g., the case  $u_0=0$, $u_L=-u_R=u$, where the energies are given by Eq.~\eqref{eq:esym}.
For complex parameters, this describes additional phase boundaries at branch cuts that signal reconfigurations of the energies in the complex plane, for instance at $d^2+u^2=0$.  This phase boundary can be crossed for a gain-loss balanced defect with $u=i\gamma$, where $\gamma$ and all other system parameters are real, which defines a parity-time (PT) symmetric setup \cite{ElG18}.
The branch cut then separates two phases where the defect energies $E_\pm$ are real (PT exact phase) or imaginary (PT broken phase). The phase boundary itself describes the conditions where  the two defect energies become degenerate in an exceptional point at $E_\pm=0$.

This behaviour is directly reflected in the effective interface theory. As we set $u_0=0$, we can work with the reduced map $\bar M$, but allow it to take complex parameters. For general $u_L$ and $u_R$, the phase boundary then occurs at $A^2=-1$.
At the phase boundary, the defect parameters $d_\pm^{(1)}$ and $\bar u_\pm^{(1)}$ of the nontrivial fixed points diverge. Entering the PT exact phase $A^2>-1$, where the energies $E_\pm$ are real, the effective defect potential $\bar u_\pm^{(1)}$ is real as well. Analogously, in the PT broken phase $A^2<-1$, where the defect energies are imaginary, the effective defect potential is imaginary, too. Examples of the renormalization flow in these two phases, as well as the associated interface states, are given in Fig.~\ref{fig:flow}.

\begin{figure}[t]
\includegraphics[width=\linewidth]{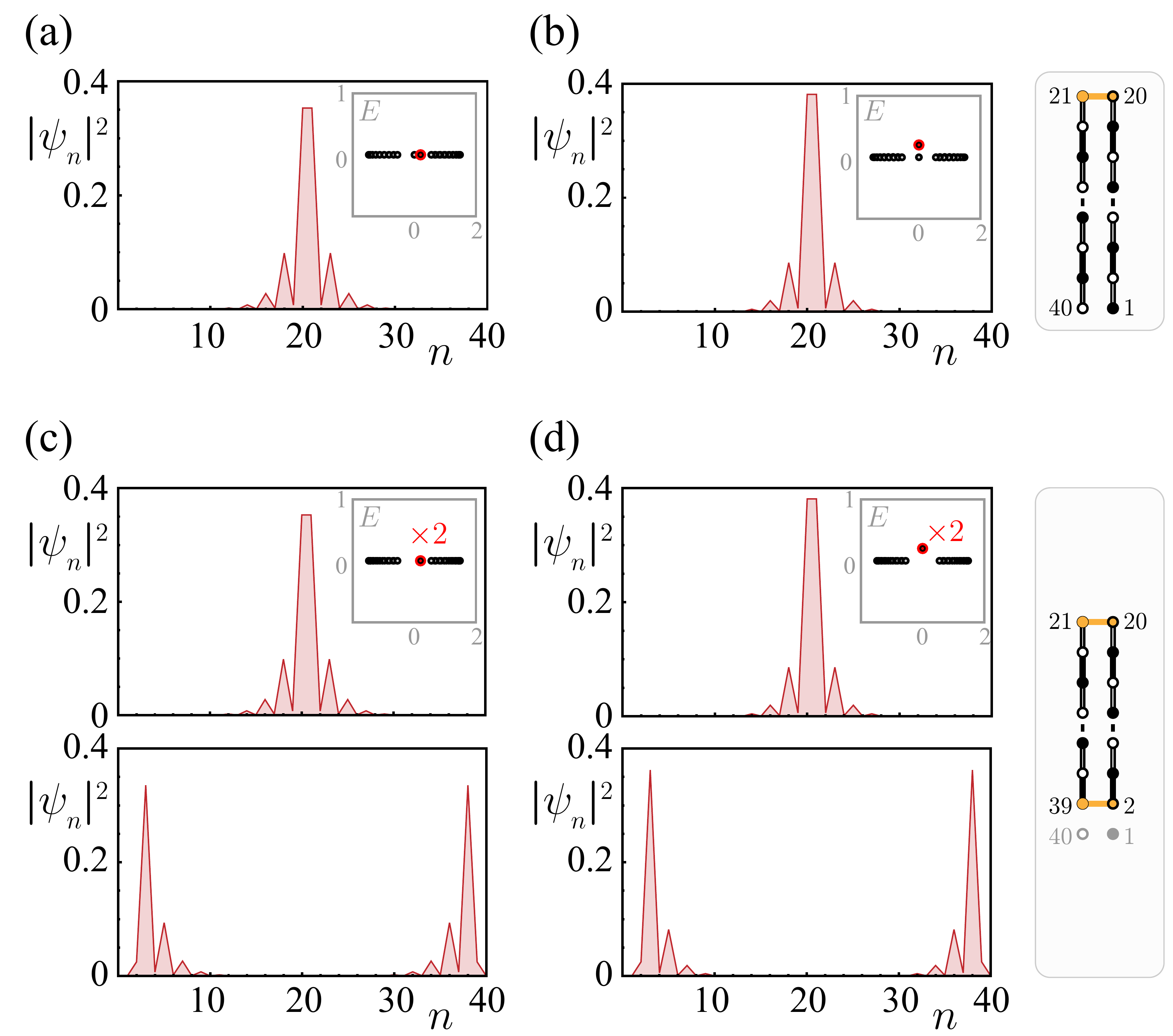}
\caption{Design of interface states with prescribed energy in (a,b) the dimer chain of Figs.~\ref{fig:dimer}, \ref{fig:flow}, as well as (c,d) a variant in which the chain is closed into a loop with a suitably matched interface at the far end (we ignore the two formally decoupled sites at the very end).
The target energy is set to (a,c) $E=0.2$ and (b,d) $E=0.2 i$, and the interface Hamiltonians are obtained from the fixed points of the reduced map $\bar M$ describing chains with $s=1/2$, $t=1$, $u_0=0$, $A=0.8 i$.
The main panels show the resulting interface states in a finite chain of 20 dimers, with the insets denoting their position in the complex energy spectrum. While the choice of $A$ nominally places the system into the PT-symmetric phase, this does not prevent the design of states with symmetry-breaking energies. With the matched interfaces, the construction becomes exact even in a finite system, and  the overall number of interface states at the target energy is maximized.
 }
\label{fig:dimerdesign}
\end{figure}

\subsubsection{Design of interface states}

As for the SSH chain, and described generally in Sec.~\ref{sec:design}, we can utilize the renormalization approach to systematically design defects that support states with desired characteristics. In particular, we can fix the energy and determine defect parameters on the stable manifold of a nontrivial fixed point, such as given by Eq.~\eqref{eq:stabman}. This is shown in Fig.~\ref{fig:dimerdesign},
where we use the fixed-point parameters at the desired energy directly, hence, fix the interface Hamiltonian to  $h=H^{(\infty)}$.
By contrasting two cases with (a) real and (b) imaginary target energy, we see that this choice is not restricted by the PT symmetry of the bulk, as the design principle automatically determines  a suitable symmetry-breaking interface where this is required.
We also show an example where the far end is equipped with a matching interface Hamiltonian $h'=H+E-h$, which here physically corresponds to a closed chain with two purposefully matched interfaces connecting the two bulk regions, as well as two isolated sites that we can remove from the description, see panels (c,d).
This results in two exact, and exactly degenerate solutions, precisely with the prescribed energy $E$, whose spatial envelop exhibits a purely exponential decay from one of the interfaces.

\section{Non-reciprocal two-legged ladder}
\label{sec:ladder}
\begin{figure}[t]
\includegraphics[width=\linewidth]{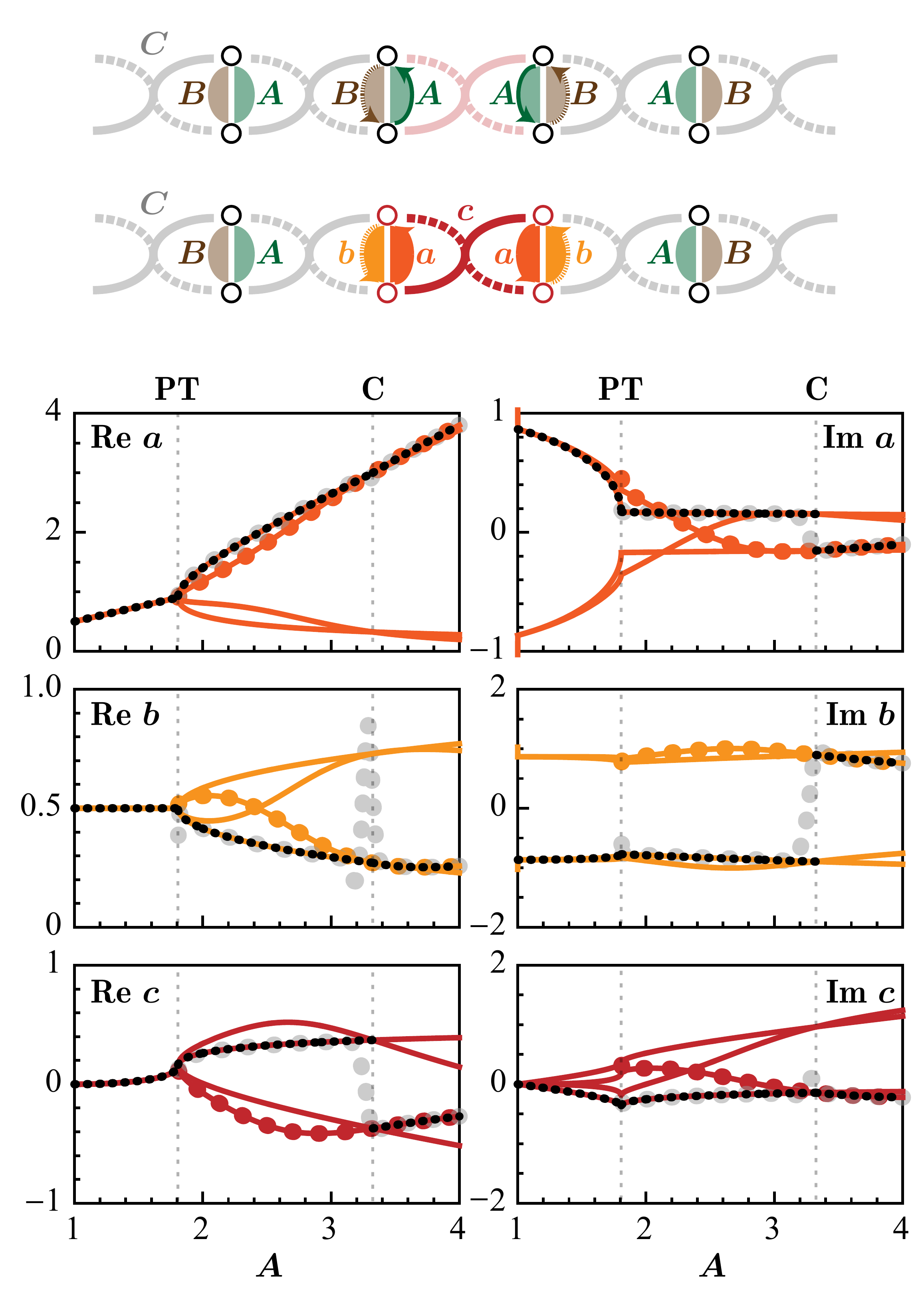}
\caption{Fixed-point selection in a non-Hermitian non-reciprocal two-legged ladder with  couplings $A\neq B,B^*$ and $C$, as sketched in the top panel. The system displays interface states without a Hermitian counterpart in a region bounded by a PT and a C symmetry-breaking phase transition.
This coincides with the conditions for which the renormalized interface parameters $a$, $b$, $c$ converge to a nontrivial fixed point,
as shown here  as a function of the bare coupling $A$ for fixed $B=C=1$ (real parts at the left, imaginary parts at the right).
The black dotted curves represent the single trivial fixed point, while the remaining solid curves represent the three nontrivial fixed points. The big colored dots show the renormalized interface parameters after 120 iterations, demonstrating that this converges to a nontrivial fixed point between the PT and C phase boundaries. For contrast, the gray dots show a generic sample trajectory, which approaches the trivial fixed point. To the left of the PT phase transition the trivial fixed point is marginally stable and represents an extended state, while the renormalization trajectories are chaotic.}
\label{fig:mpsconvergence}
\end{figure}

In this section we apply the renormalization approach to a quasi-one-dimensional system with non-reciprocal non-Hermitian couplings. These aspects are combined in the system sketched in Fig.~\ref{fig:mpsconvergence}, consisting of a two-legged ladder with asymmetrical couplings $A\neq B,B^*$, with an interface that separates regions where the role of these couplings are interchanged.  This system can support defect states that have no Hermitian counterpart, which emerge at phase boundaries that signal the breaking of a PT symmetry or a non-Hermitian charge conjugation (C) symmetry  \cite{Mal15}.
To analyze these features in the general approach, we set the bulk data to
\begin{align}
H=\begin{pmatrix}
    0 & A &  0 & 0 \\
    B & 0  &  0 & 0\\
    0 & 0 & 0 & A \\
    0 & 0 & B & 0 \\
  \end{pmatrix},
  \quad V=W^T=
  \begin{pmatrix}
    0 & C & 0 & 0 \\
    C & 0 & 0 & 0 \\
    0 & 0 & 0 & C \\
    0 & 0 & C & 0
  \end{pmatrix}.
\end{align}
and initialize the renormalization map \eqref{eq:genmap}  with
\begin{align}
H^{(1)}=\begin{pmatrix}
    0 & A &  C & 0 \\
    B & 0  &  0 & C\\
    C & 0 & 0 & A \\
    0 & C & B & 0 \\
  \end{pmatrix}.
\end{align}
Because of the symmetries in this construction, the successively renormalized Hamiltonian retains the structure
\begin{align}
H^{(n)}=
\frac{1}{2}
\begin{pmatrix}
    c_+ & a_+  \\
    b_+ & c_+  \\
  \end{pmatrix}\otimes\begin{pmatrix}
    1 & 1  \\
    1 & 1  \\
  \end{pmatrix}
+
  \frac{1}{2}
\begin{pmatrix}
    c_- & a_-  \\
    b_- & c_-  \\
  \end{pmatrix}\otimes\begin{pmatrix}
    1 & -1  \\
    -1 & 1  \\
  \end{pmatrix}
,
\end{align}
which corresponds to the sectors of spatially symmetric and antisymmetric wave functions.
The defect parameters therefore separate into two sets, which both obey the same map
\begin{align}
a'&=A+\frac{b C^2}{(c-E)^2-a b},
\nonumber\\
b'&=B+\frac{a C^2}{(c -E)^2-a b},
\nonumber\\
c'&=\frac{(c -E)C^2}{a b-(c-E)^2}.
\label{eq:renormmps}
\end{align}
However, both parameter sets take different initial conditions $(a_\pm,b_\pm,c_\pm)=(A,B,\pm C)$.

\begin{figure*}[t]
\includegraphics[width=0.9\linewidth]{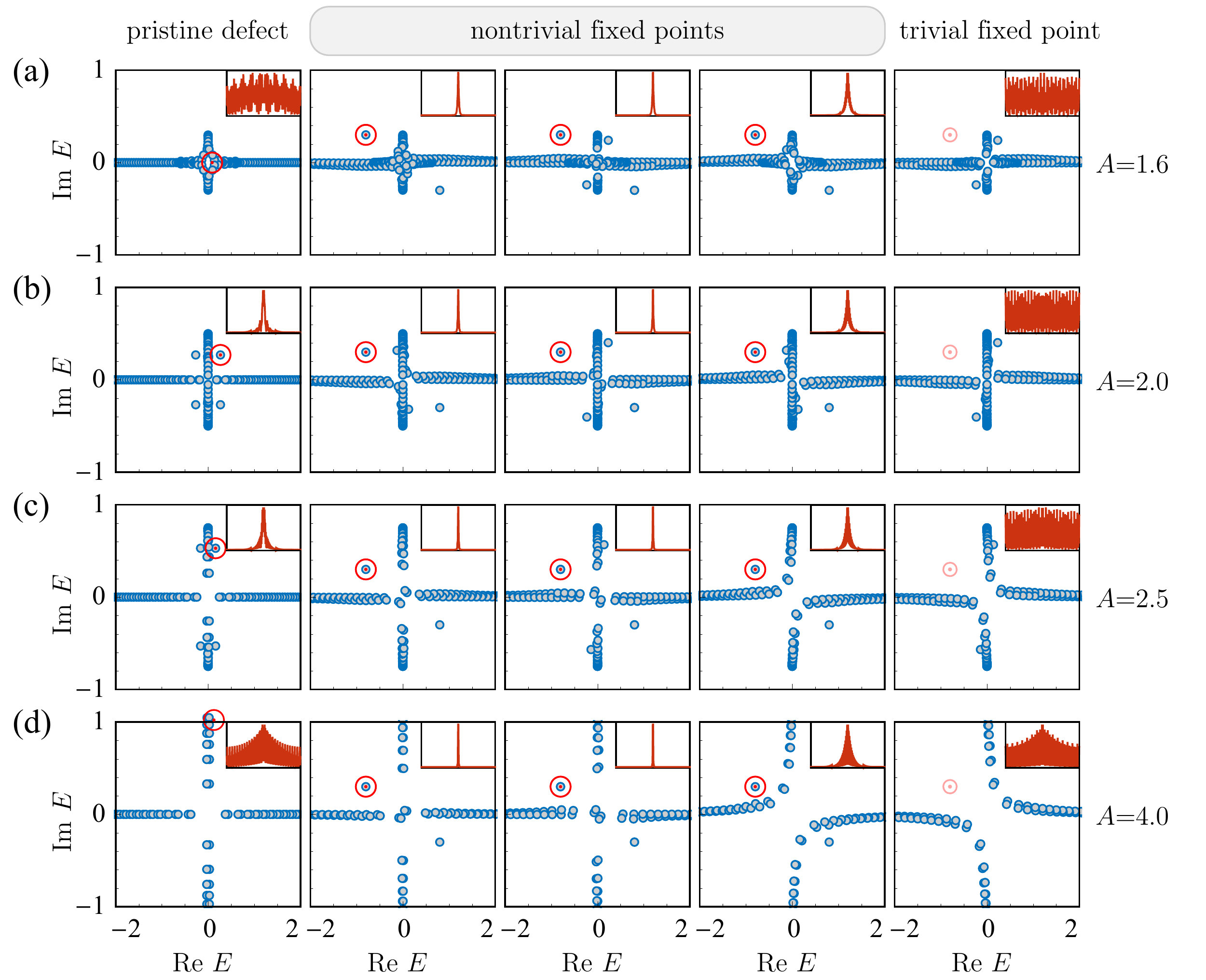}
\caption{Design of interface states with a predetermined energy in the nonreciprocal two-legged ladder, as illustrated by complex energy spectra (main panels) and intensity profiles (insets) in systems of 120 resonators with bulk couplings $B=C=1$ and  (a) $A=1.6$, (b) $A=2$, (c) $A=2.5$, and (d) $A=4$.
For comparison, the left column shows  the system in the pristine interface configuration of Fig.~\ref{fig:mpsconvergence}, which supports interface states only in panels (b) and (c) where $A$ lies between in the PT and C phase boundaries. The energy of these states then follows from Eq.~\eqref{eq:mpsquant}, as indicated by a red dot.
In the remaining columns, the interface parameters are obtained from the fixed points \eqref{eq:mpsfix} of the renormalization map \eqref{eq:renormmps}, where the target energy  (again indicated by a red mark) is set to $E=-0.8+0.3 i$.
In all cases, an interface state at the desired energy then exists provided the parameters are taken from one of the three nontrivial fixed points (second to fourth column).}
\label{fig:mpsdesign}
\end{figure*}

For any given energy, there are four fixed points, which can be written as
\begin{align}
X^{(\pm,\pm)}:\quad&c=\frac{E}{2}\left(1\pm\sqrt{y\pm\sqrt{y^2-(A+B)^2/ x}}\right),
\nonumber\\
&a=\frac{E-c}{E(E-2c)}[A(E-c)+Bc],
\nonumber\\
&b=\frac{E-c}{E(E-2c)}[B(E-c)+Ac],
\label{eq:mpsfix}
\end{align}
where
\begin{align}
x&=(A-B)^2+4 E^2,
\\
y&=(A^2+B^2+2 E^2-8 C^2)/x.
\end{align}
A stable manifold of these crosses the initial bare interface parameters $(A,B,\pm C)$ when the energy fulfills
\begin{equation}
(E \mp 2 C)^2E-  AB E = \pm (A - B)^2 C/2,
\label{eq:mpsquant}
\end{equation}
which then  represents the quantization condition of states with symmetric or antisymmetric wave functions.
The PT and C symmetry-breaking  phase transitions occur when fixed points change their stability,
as illustrated in Fig.~\ref{fig:mpsconvergence}. There, we identify the trivial fixed point by initializing a trajectory at a generic initial condition. We then see that in the region between the PT and C phase boundaries, where interface states exist, the initial conditions converge to a different, non-trivial fixed point. At the PT phase transition, the trivial fixed point becomes marginally stable, so that beyond this boundary trajectories do not converge. Instead of interface states, the system then supports extended states at the predicted energy. Beyond the C phase transition, the renormalization trajectory converges to a stable fixed point, and the interface state disappears as it is no longer normalizable.

We note that the renormalization flow can explicitly introduce PT and C symmetry breaking, as the couplings $a,b,c$ depart from their bulk values.
The quantization condition for such a general defect is \emph{much} more complicated than Eq.~\eqref{eq:mpsquant}
(written out it fills several pages), but
it can be checked that it is invariant under the renormalization map \eqref{eq:renormmps}. Therefore, it can be efficiently analyzed in the developed approach.
In particular, using the fixed points of this map, we can again design defects that support defect states with a predetermined energy, even for bulk values where no such states exist in the original setting.
For this, we insert the bulk and parameters and target energy into Eq.~\eqref{eq:mpsfix},
which gives us a choice of microscopic defect parameters.
In the flow, these parameter are not connected to a trajectory passing through $(a,b,c)=(A,B,\pm C)$, but explore different region in the defect parameter space.

With this principle, we can create defects state whose energy can be freely placed in the complex plane. This is illustrated in Fig.~\ref{fig:mpsdesign} for a system of 120 resonators. We see that the three non-trivial fixed points induce localized defect states with the desired energy, irrespective of the phase of the system with pristine defect parameters. The trivial fixed point does not induce such a state, which provide another method to identify it. Furthermore, the three non-trivial fixed points have different stabilities, which directly determines how strongly the induced states are localized (see insets). We also note that the energy of all other states shows a dependence on the choice of the fixed point. This is particularly the case for initially already localized states, while the bands of extended states show a small but systematic splitting that diminishes slowly with increasing system size.
This contrasts with the desired interface state, whose energy and intensity profile is already well established at the displayed system size.

\begin{figure}[t]
\includegraphics[width=\linewidth]{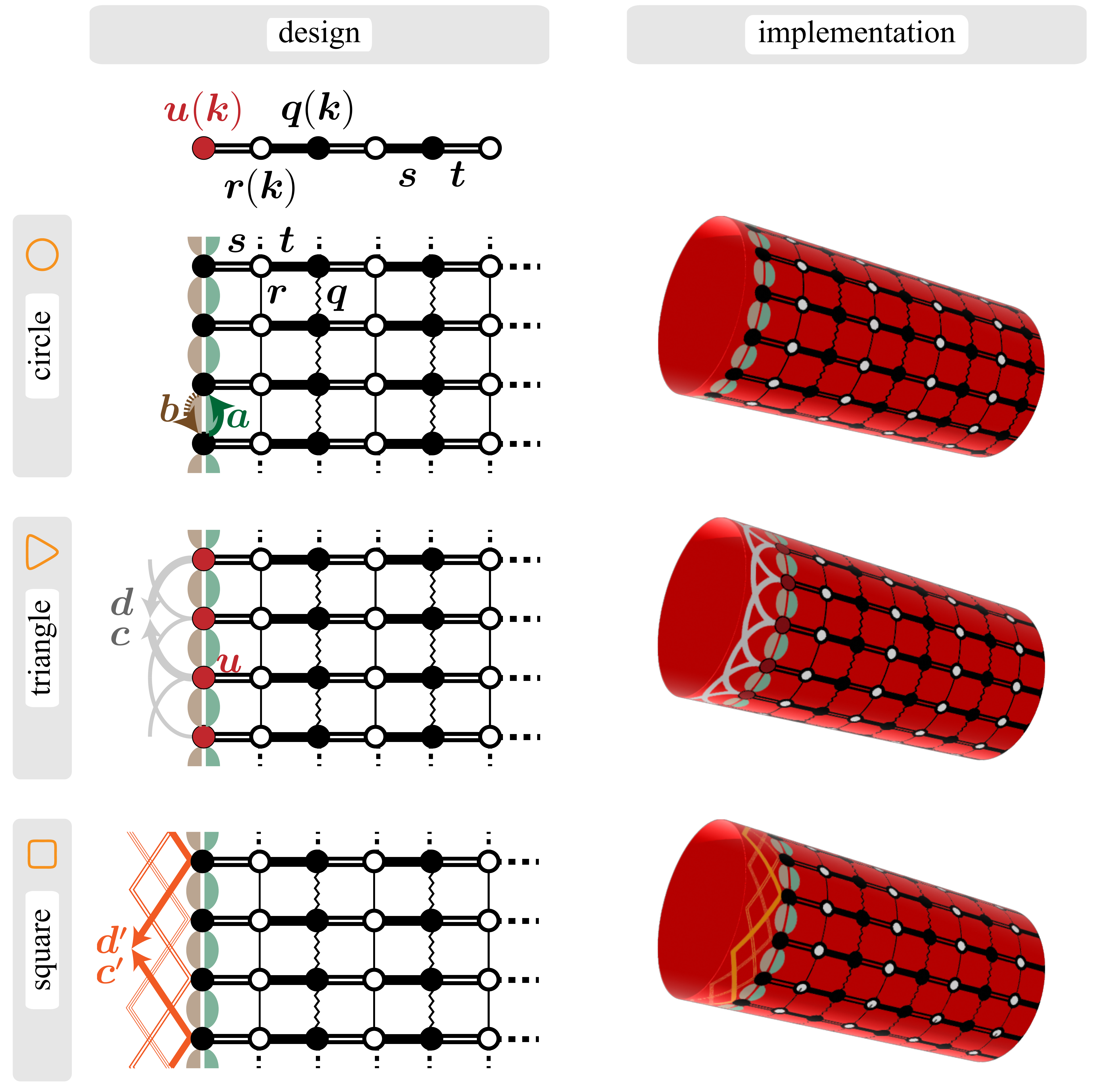}
\caption{Design principles for edge dispersion relations in a quasi-one dimensional system based on coupled SSH chains.
The target dispersions approximate geometric shapes as indicated in the legend, where the yellow curves  are obtained from Eq.~\eqref{eq:dispapprox}.
The left panels show the lattice design with bulk couplings $q$, $r$,  $s$, $t$ and edges configured in accordance to the target edge dispersion.
For a circular target dispersion (top row), this utilizes nonreciprocal nearest-neighbor couplings $a$ and $b$.
For a target dispersion approximating a triangle (middle row), these couplings are supplemented by an onsite potential $u$ and nonreciprocal next-nearest-neighbor couplings $c$ and $d$. For the square (bottom row), we use the couplings
$a$ and $b$ as well as third-nearest-neighbor couplings $c'$ and $d'$. In all cases, the translational invariance along the edge direction can be used to transform the system into a chain with parameters depending on the transverse wavenumber $k$, as indicated at the very top.
The right panels illustrate the implementation as a finite system on a cylinder. For results see Fig.~\ref{fig:sshq1dall}.
}
\label{fig:sshq1dsketch}
\end{figure}

\begin{figure*}[t]
\includegraphics[width=0.9\linewidth]{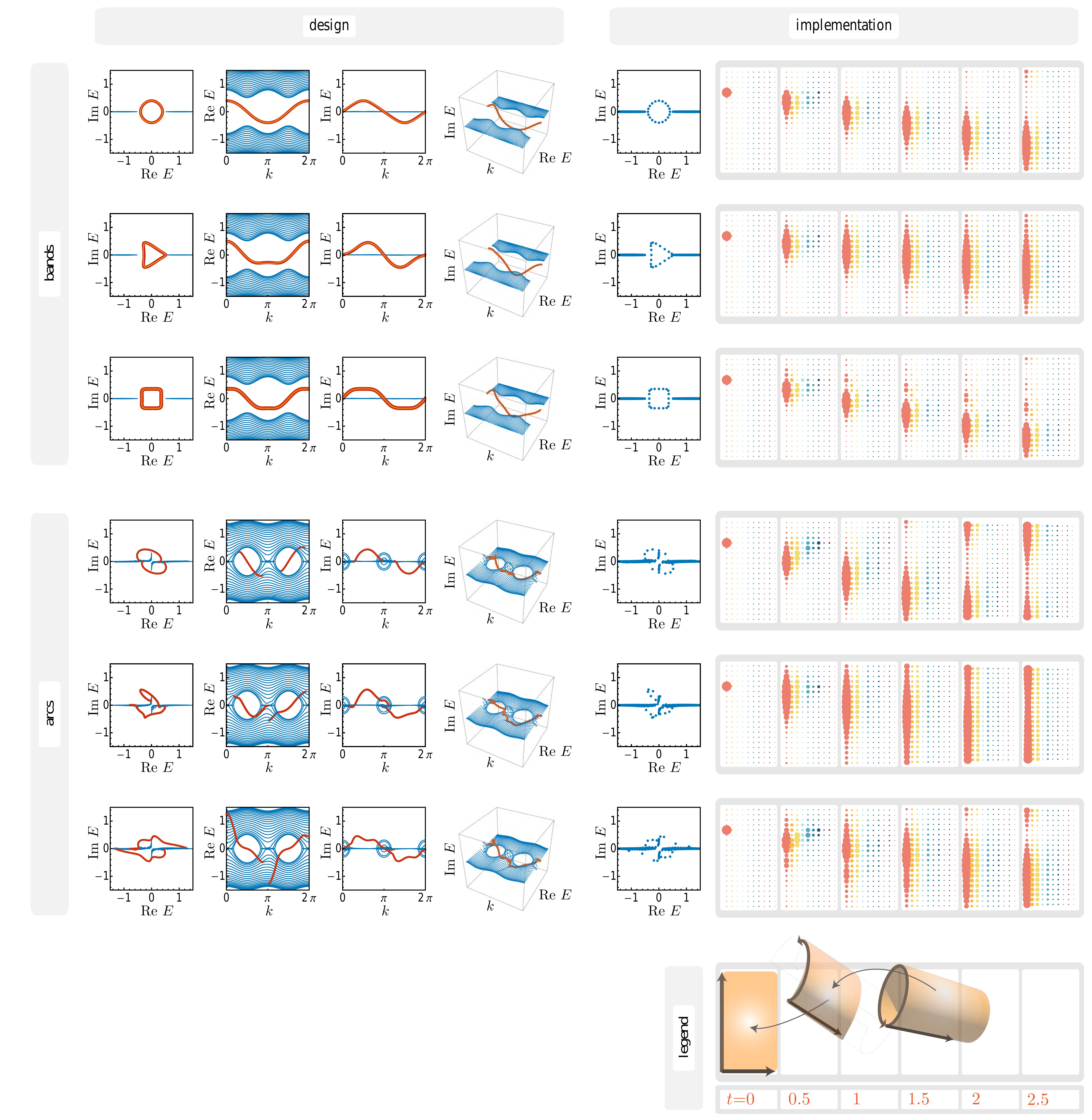}
\caption{Geometric edge dispersion band and associated arcs obtained from the design in Fig.~\ref{fig:sshq1dsketch}.
As shown in the top half of the figure, edge bands are obtained when the target dispersion is placed into the bulk gap of the system. This is here realized for a system with bulk couplings $s=1/2$, $t=1$, $r=-q=0.3$, and target edge dispersions \eqref{eq:dispapprox} with
$E=0.4$ (all cases), $E_0'=0.1$ (triangle), $E_0'=-0.05$ (square). The design panels on the left confirm these bands in the effective $k$-dependent representation of the system, where the realized edge dispersion (red) corresponds well to the target dispersion (overlayed in orange), while the bulk states are shown in blue.
The implementation panels on the right show that the desired edge states also appear in a finite system (20 chains of length 41), which then supports unidirectional edge transport according to the different lifetimes of the branches with a positive or negative group velocity (time snapshots of the edge region of the cylinder, with the time steps and coordinate system specified in the legend).
As shown in the bottom half of the figure,
arcs are obtained when the target dispersion (of the same shape as in the top panels, and hence omitted) intersects with the bulk bands, which we here obtain by changing the bulk couplings between the chains to $r=-q=0.3i$.
}
\label{fig:sshq1dall}
\end{figure*}

\section{Design of symmetry-breaking edge  bands and arcs}
\label{sec:bandsandarcs}

As our final example, we consider interface states along a one-dimensional edge of a two-dimensional system, which can be interpreted as a quasi-one-dimensional arrangement of SSH chains joined up with additional alternating bulk couplings $q$ and $r$
as shown in Fig.~\ref{fig:sshq1dsketch}. Our goal is to functionalize the edge to produce a desired edge-state dispersion relation $E(k)$, prescribing a loop or arc in the complex energy plane as a function of the transverse wavenumber $k$. For this, we utilize a variety of periodic edge-coupling configurations as illustrated in the left panels, and further detailed below. For the final implementation we consider a finite system placed onto a cylinder, with periodic boundary conditions along the direction of the edge, as shown in the right panels.

To set up the general approach, we exploit the general periodicity of the system into the direction along the edge, so that all solutions of the wave equation can be parameterized by $k$.
In the bulk, this determines the matrices
\begin{align}
H=\begin{pmatrix}
    2 q\cos k  & s \\
    s & 2 r\cos k\\
  \end{pmatrix},
  \\
V=W^T=\begin{pmatrix}
    0  & 0 \\
    t & 0\\
  \end{pmatrix},
\end{align}
where we note that the couplings $q$ and $r$ break the chiral symmetry.
At the edge, the effective potential $u$ becomes $k$-dependent as well,
\begin{equation}
H^{(1)}=\begin{pmatrix}
    u(k)  & s \\
    s & 2 r\cos k\\
  \end{pmatrix}
\end{equation}
where the specific form of
\begin{equation}
u(k)=\sum_n U_ne^{nik}
\end{equation}
depends on the edge couplings $U_n$ connecting sites to their $n$th nearest neighbor. We allow these couplings to be non-Hermitian and nonreciprocal, so that in general $U_n\neq U_{-n}^*, U_{-n}$.

Our strategy is to specify a target edge dispersion $E(k)$ within the bulk band gap, introduce edge couplings $U_n$ that respect the symmetries of this dispersion, and then use a minimal number of regularly spaced values $k=l\Delta k$ ($l=0,1,2,\ldots$) to
determine the values of these couplings from the fixed points of the renormalization map.

Specifically, we consider target dispersions of the form
\begin{align}
E(k)&=E_0e^{ik}&\mbox{circle,}\nonumber\\
E(k)&=E_0e^{ik}+E_0'e^{-2ik}&\mbox{triangle,}\nonumber\\
E(k)&=E_0e^{ik}+E_0'e^{-3ik}&\mbox{square},
\label{eq:dispapprox}
\end{align}
with parameters chosen to approximate the specified geometric shapes.
For the circle, we utilize the two couplings $U_{-1}=a$ and $U_{1}=b$, and determine their values from the fixed points at $k=0$ and $k=\pi/2$ (this automatically also enforces agreement at $k=\pi$ and $k=3\pi/2$). For the triangle we supplement the couplings $a$ and $b$ by couplings $U_{0}=u$, $U_{-2}=c$, $U_{2}=d$, and determine their values at multiples of $\Delta k=2\pi/5$. For the square, we use the couplings $a$ and $b$ as well as $U_{-3}=c'$, $U_{3}=d'$, and determine their values at multiples of $\Delta k=\pi/4$. This results in edge couplings that depend on the chosen values of $E_0$ and $E_0'$  as well as the bulk parameters of the  system.

The results from this procedure are shown in Fig.~\ref{fig:sshq1dall}. In the upper panels (`bands'), we set $s=1/2$, $t=1$, $r=-q=0.3$, for which the bulk band structure is real and gapped. For the target dispersions, the parameters are chosen as $E=0.4$ (all cases) and $E_0'=0.1$ (triangle), $E_0'=-0.05$ (square), approximating the desired geometric shapes within the gap. These shapes are then clearly displayed in the actual $k$-dependent band structure of the designed system, as shown in the left panels for chains of length $41$. This also holds for the final implementation of the system on a cylinder with 20 chains, for which results are shown in the right panels.

The right panels also confirm a key feature of the designed edge dispersions, namely that they equip the dispersion branches of opposite group velocity $v=\mathrm{Re}\,dE/dk$ with different life times $\propto(\mathrm{-Im}\,E)^{-1}$. For the chosen dispersions, this results in wave components propagating in the negative edge direction to have a longer life time than those propagating in the positive direction. As confirmed by the snapshots, a wavepacket that is initially localized on a single edge site then evolves into a wavepacket that propagates in the negative direction around the edge of the cylinder.

As shown in the lower panels of Fig.~\ref{fig:sshq1dall} (`arcs'), the edge bands can be transformed into arcs when we design them to interact with the bulk band structure. Here, this is enforced by setting $r=-q=0.3 i$, so that the bulk dispersion displays branch points connecting real and imaginary dispersion branches. The resulting edge arcs are present both in the $k$-dependent dispersion in the left panels and in the implementation on the cylinder in the right panels. The resulting wave propagation still displays filtering into a preferred propagation direction, but with increased broadening and leakage into the bulk.

\section{Nonlinear effects from saturable gain}
\label{sec:nonlin}

Topological and non-Hermitian systems that incorporate nonlinear effects attract attention as they
offer highly promising routes to achieve nonreciprocity and active tunability \cite{Smirnova2020}. In addition, many desired non-Hermitian effects, such as PT-symmetry, topological lasing, and the non-Hermitian skin effect, essentially rely on active components, which are intrinsically nonlinear \cite{ElG18,Oza19,Ota20}.
In the context of the full coupled-mode equations, these nonlinearities significantly change the  nature of the problem. Instead of eigenstates with a complex energy, which mathematically are only well defined in linear settings, one is then interested in  finding self-consistent stationary solutions
with effectively real energies (as only for such energies the intensities are stationary). In the space of the original coupled-mode equations, this corresponds to a prohibitively large set of nonlinear equations. A standard approach is to instead propagate the system from generic initial condition, which requires numerically expensive time integration, and again has to sample over a  high-dimensional space of initial conditions. Furthermore, amongst the obtained stationary operation modes, edge and interface state have then to be selected by their mode profile, which can be a challenging task in itself.

We therefore here describe how to incorporate nonlinearities into the renormalization approach. Indeed, because of its
intrinsically nonlinear nature, this approach naturally extends to such settings, and then offers a conceptually consistent and extremely efficient method that continues to operate in a very small parameter space.
For this, we amend the renormalization parameters by the intensities within a unit cell, which are automatically embedded in the described recursion formalism, as given by equations
\eqref{eq:recrel}. Edge and interface states are then again characterized by the convergence to an unstable fixed point. Furthermore, as this automatically guarantees that the intensities decay into the bulk, the relevant fixed points can be obtained from the low-intensity approximation. At the same time, the full intensity profile  throughout the whole system is accounted for completely by the renormalization trajectory itself, which is obtained efficiently from the renormalization map.

\begin{figure}[t]
\includegraphics[width=\linewidth]{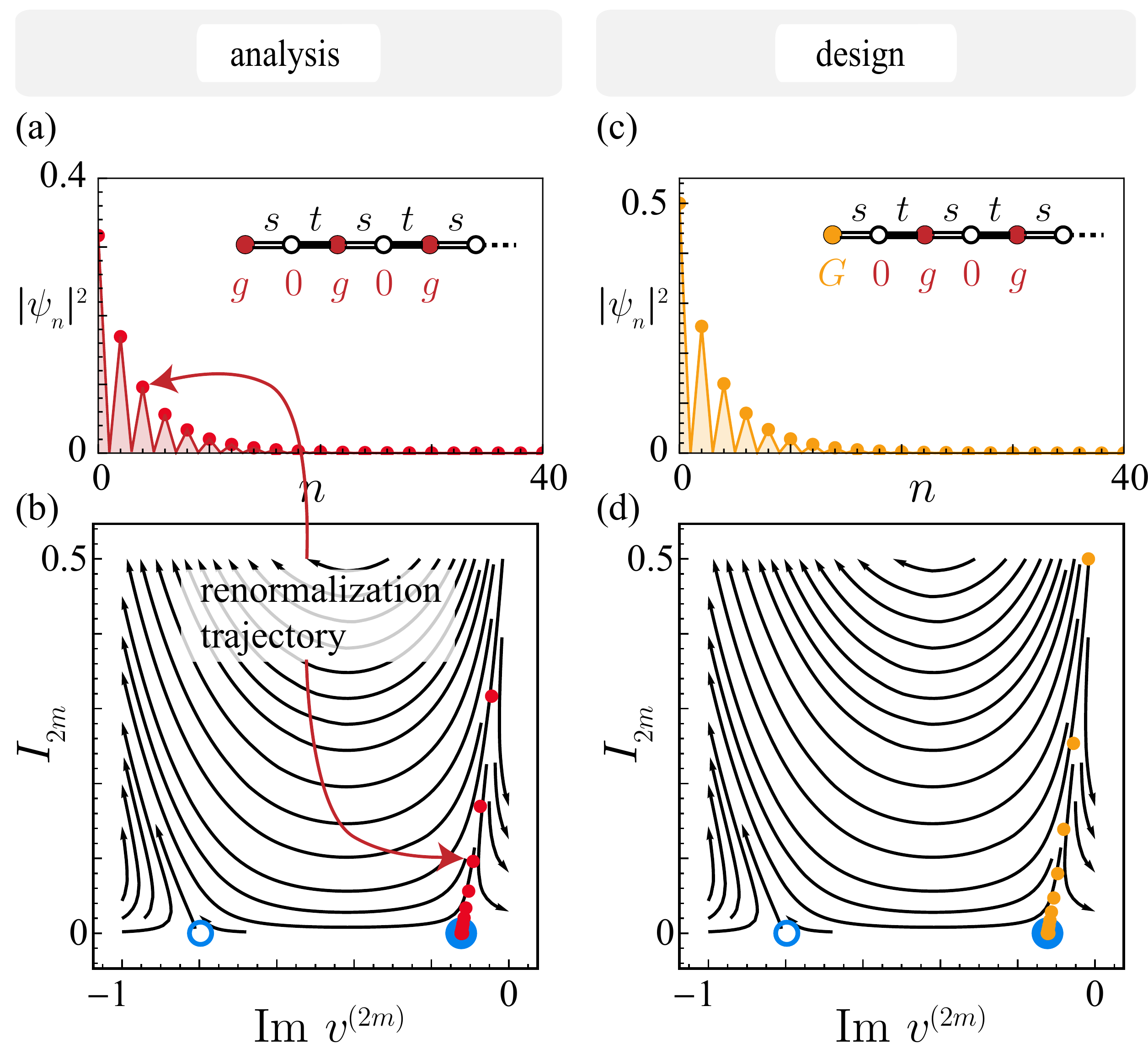}
\caption{Analysis and design of edge states in a nonlinear dimer chain with staggered couplings $s=0.7$ and $t=1$ and nonlinear distributed gain and loss as specified in Eq.~\eqref{eq:nonlingain}, where $g=0.6$ on the even sites, $g=0$ on the odd sites, and $\gamma=0.5$, $S=1$. This configuration realizes a dynamically stable stationary edge state, which can be obtained numerically by propagation out of a suitable initial condition. In panel (a), this intensity profile is shown as the shaded curve, obtained by Crank-Nicolson propagation over 50000 dimensionless time steps $\delta t=0.01$ in a system of 100 sites, where convergence is verified from the self-consistency condition of the instantaneous Hamiltonian.
The red dots in this panel are the intensity profile obtained from the corresponding renormalization trajectory, Eq.~\eqref{eq:nonlinrenomdimer},  which now not only accounts for the renormalised effective potential $v^{(n)}$, but also includes the intensity $I_{n}$ on each site.
Panel (b) shows the corresponding renormalization flow for the iteration over one dimer unit cell. The red dots again show the renormalization trajectory of the edge state, verifying that it converges to a low-intensity fixed point (blue dot).
The self-consistent initial condition $I_0=0.3167...$ of this trajectory is obtained highly
efficiently by an exponentially quickly converging bisection search, which places it onto the stable manifold of this fixed point.
In panels (c) and (d), the same renormalization flow is used to design a system with modified pumping strength $G$ at the edge, such that it supports a stable edge state with edge intensity $I_0=0.5$. This is realized by setting the edge gain parameter to $G=0.7243...$,
with this value again determined efficiently by bisection. Again, we obtain precise agreement with the numerically obtained intensity profile [dots versus shaded curve in panel (c)].}
\label{fig:nonlin}
\end{figure}

In the following, we demonstrate the remarkable efficiency, consistency, and practicability of this procedure for the paradigmatic model of a dimer-chain with distributed saturable gain and loss (see insets in Fig.~\ref{fig:nonlin}) \cite{Malzard2018,Lang18,Wong2023}, which has been realized in several laser experiments \cite{St-Jean2017,Zha18,Par18,Ota2018,Whittaker2019}. A motivating feature for the design of these systems is that they can support symmetry-protected edge states even in the presence of non-linear gain saturation.
In the coupled-mode equations
\begin{equation}
t_n\psi_{n-1}+[v_n(I_n)-E]\psi_n+t_{n+1}\psi_{n+1}=0
\end{equation}
of the chain,
this saturation is modelled by  nonlinear onsite terms
\begin{equation}
\label{eq:nonlingain}
v_n(I_n)=\frac{i g_n}{1+S_n I_n}-i\gamma_n
\end{equation}
that depend on the intensity $I_n=|\psi_n|^2$, where $g_n$ characterizes the gain on the component, $S_n$ determines the nonlinear saturation, and $\gamma_n$ accounts for additional background losses.
The gain and loss are distributed to select the edge state of the linear SSH model discussed in Sec.~\ref{sec:ssh}, which arises from the dimerized coupling pattern
$t_n=s$ for odd $n=2m+1$ and $t_n=t$ for even $n=2m$, along a semi-infinite system with site index $n\geq 0$ (the index $m\geq 0$  enumerates the dimers.)
The desired distributed gain is obtained by setting $g_n\equiv g$ on the sublattice with even $n$, while $g_n=0$ on the odd-numbered sites. As several laser implementations utilize additional pumping at the edge, we allow $g_0\equiv G$ to take a different value. Furthermore, we set $S_n=1$ to fix the overall intensity scale, and set the background losses to a common value $\gamma_n\equiv \gamma$ on all sites.

The  phase space of this model has been studied before within conventional, computationally intensive approaches, including machine-learning based algorithms to discriminate the edge states \cite{Wong2023}.
Their symmetry-protection arises from a nonlinear extension of charge-conjugation symmetry, which applies at self-consistent energy $E=0$ of the instantaneous Hamiltonian \cite{Malzard2018a}.
An example of such a numerically obtained edge state is shown as a shaded intensity profile in Fig.~\ref{fig:nonlin}(a), with the model parameters given in the caption.

Following the described nonlinear renormalization procedure, we can instead address this problem by a renormalization map
that combines Eqs.~\eqref{eq:genmap} for the system parameters
with Eq.~\eqref{eq:recrel} for the wavefunctions.
In the present model, this then results in a nonlinear recursion relation
for the intensity $I_n$ and the effective onsite potential  $v^{(n)}$,
\begin{align}
I_{2m+1}&=\left|\frac{E-v^{(2m)}}{s}\right|^2I_{2m},
\nonumber
\\
v^{(2m+1)}&=\frac{s^2}{E-v^{(2m)}}-i\gamma,
\nonumber
\\
I_{2m+2}&=\left|\frac{E-v^{(2m+1)}}{t}\right|^2I_{2m+1},
\nonumber
\\
v^{(2m+2)}&=\frac{t^2}{E-v^{(2m+1)}}+\frac{ig}{1+I_{2m+2}}-i\gamma,
\label{eq:nonlinrenomdimer}
\end{align}
where we explicitly spelled out all steps for the iteration over a complete dimer unit cell.
In the linear limit of small intensities $I_{n}$, the equations for the onsite potential close, so that they can be determined from
Eqs.~\eqref{eq:genmap}. The additional equations for the intensities then enforce self-consistency of the solution along the whole system.
The renormalization trajectory is then initialized by the edge intensity $I_0$, which also fixes the bare nonlinear edge potential
\begin{equation}
v^{(0)}=\frac{iG}{1+I_{0}}-i\gamma.
\end{equation}

We focus on the symmetry point $E=0$, where the renormalized potential $v^{(m)}$ is purely imaginary, which reflects the above-mentioned nonlinear extension of charge-conjugation symmetry.   Figure \ref{fig:nonlin}(b) shows the corresponding renormalization flow, along with a renormalization trajectory that converges to the unstable fixed point with $I_\infty=0$,
\begin{align}
v^{(\infty)}=&i \frac{s^2-t^2+(g-\gamma)\gamma}{2\gamma}
\nonumber\\
&+i \frac{\sqrt{((s - t)^2 - (g-\gamma)\gamma) ((s + t)^2 -
   (g-\gamma)\gamma)}}{2\gamma}
   .
   \end{align}
This immediately results in a comprehensive characterization of the edge states in the system.
Firstly, the requirement that this fixed point is purely imaginary directly provides previously unknown analytical conditions for the existence of this state.
Secondly, as highlighted in the Figs.~\ref{fig:nonlin}(a,b), the renormalization trajectory converging to this fixed point precisely recovers the intensity pattern of the numerically obtained edge state.
Thirdly, the initial condition of this trajectory can be determined very efficiently by a simple bisection approach of $I_0$, which converges exponentially quickly and in practice is carried out instantaneously, in stark contrast to the previous approaches to this problem.
Finally, as before, the same approach can also be utilized to design interfaces that support desired states. This is illustrated in Fig.~\ref{fig:nonlin}(c,d), where we determine the edge gain parameter $G$ so that the system supports an edge state with  a prescribed edge intensity $I_0$. Again, the required value of $G$ can be obtained efficiently by bisection.
Therefore, the renormalization approach to this problem is practical and efficient, and delivers direct quantitative information and original conceptual insights into the edge states of the system.

\section{Discussion and conclusions}
\label{sec:conclusions}

In summary, we have presented an efficient real-space renormalization approach that provides a universal description of interface states over a wide range of Hermitian and non-Hermitian models, irrespective of whether they obey a traditional bulk-boundary principle or not.
The approach rests on the map \eqref{eq:genmap}, which is completely defined by the bulk data, and  condenses the relevant interface parameters into a typically very low-dimensional space. As shown in Sec.~\ref{sec:stabquant}, the stability of the fixed points in the map is directly related to the asymptotic behaviour of the interface states, so that their quantization condition identically maps onto the question of convergence to an unstable fixed point. In turn, these fixed points and their stable manifolds can then be used to design interfaces with desired properties.

As shown in the extensive examples, this approach can be  applied to models with symmetry-breaking potentials, non-Hermitian phase transitions,  non-reciprocal couplings, and finite-dimensional edge dispersions, with the main requirement being the periodicity in the bulk. The approach captures the rich phenomenology of interface states, including cases where they can be interpreted as topological states or their deformations, or where they are induced by the interface itself, and links these scenarios to structural aspects of fixed points and manifolds in a nonlinear system, which can reconfigure, e.g., in bifurcations.

There is an abundant set of further paradigmatic models to which this approach can be directly applied to, including
systems with flat bands \cite{Leykam18} or higher-order topological states \cite{Schindler18} such as based on the Lieb \cite{GuzmanSilva2014,Poli2017} or Kagome \cite{Kun18a,Ezawa18} lattice. With straight interfaces aligned into different directions,
 the discussed approach results in a systematic dimensional reduction but the resulting effective models are expected to be highly complex. Furthermore, the approach can be applied to
topological paradigms such as the Haldane \cite{Hal88} or Kane-Mele model \cite{Kane05}, or the wide classes of square-root topological insulators \cite{Ark17}.  Interesting is also the application to nonsymmorphic systems \cite{Shiozaki14} such as shrunken and expanded honeycomb lattices \cite{Wu15}, where interfaces necessarily break the topologically relevant bulk symmetries.
In all these systems, new effects can then be induced by the general interface design principles.
Technologically important is the fact that this allows to place the interface energies freely in the complex plane, so that their physical signatures can be purposefully enhanced or suppressed, which leads to direct applications in the design of non-Hermitian sensors, amplifiers, lasers, and filters \cite{Sch13,St-Jean2017,Zha18,Par18,Wiersig20,Sch20,McDonald2020,Wanjura2020,Budich2020,Price2022}. Here we have already demonstrated one such filtering mechanism in the design of geometric edge dispersions that result in unidirectional transport.

Being based on a coupled-mode tight-binding description, the analysis and design principles can be applied to a wide range of physical platforms, including the comprehensive set of photonic resonator and waveguide arrays and patterned semiconductor structures reviewed in Refs.~\cite{Oza19,Price2022}, suitably structured excitonic systems \cite{Mangussi2020},
as well as analogous mechanical \cite{ghatak2020} and electronic \cite{Helbig2020} designs.
The emphasis on these models therefore reflects the nature of the paradigmatic topological and non-Hermitian models presently utilized both in theory and for guidance in the design of experiments in transferrable settings.
However, just as the recursive Green's function method for transport calculations \cite{datta1997electronic}, the approach also embeds naturally into microscopic modelling frameworks beyond the coupled-mode approximation. This is particularly relevant for photonic crystals modelled on the level of Maxwell's equations.
We therefore note that the formalism can be incorporated into recursive-Greens-functions based real-space discretization approaches of Maxwell's equations \cite{Rahachou2005}, and also links to the boundary-integral approach, which defines the state-of-the art for dielectric microresonator modelling  \cite{Wiersig2003}. Besides their efficiency, a particular strength of these approaches is that they consistently account for radiative boundary conditions via surface Green's functions. In the presented framework, these surface Green's functions then define the initial conditions for the renormalization. The proposed formalism can therefore directly capitalize on the intrinsic strengths of these microscopic modelling approaches.

Indeed, a highly advantageous feature of the presented framework is that it naturally ties into generalized physical settings.
In Sec.~\ref{sec:nonlin}, we already described the extension to nonlinear systems.
Analogously, it is also worthwhile to consider how the approach can be embedded, either numerically or conceptually, into descriptions of interacting systems, for instance self-consistently on a nonlinear mean-field level, or as input into a configuration-interaction description. Further extensions could consider the role of bulk disorder, mathematical connections to bulk symmetry classifications and topological invariants, and systems with periodicity in synthetic dimensions  \cite{Ozawa2019}.

Many of these applications and extensions require to consider the role of bulk states in the presented renormalization approach, and therefore we close with a few remarks on these.
Applied to extended states in systems with an interface, biorthogonal quantization theory implies that these correspond to superpositions of states with different complex propagation factors, whose right and left versions can be combined to display a quasistationary beating pattern deep in the bulk \cite{Edvardsson2020}. Furthermore, in the most general case where the system is non-Hermitian and nonreciprocal, the resulting energy spectrum of these states can be drastically different from the conventional Bloch band structure of the infinite periodic system. On the other hand, for Hermitian systems, the energy ranges of extended states in both physical setting are in close correspondence, which provides the backdrop of the bulk-boundary correspondence in systems with additional topological symmetries \cite{Asboth2016}.

We now observe that in the language of the renormalization approach of this work, the recursion relations
Eq.~\eqref{eq:recrel} imply that the quasistationary patterns correspond to initial conditions on specific manifolds from which the renormalization map does not converge.
In this work, we already encounter this behaviour in two specific examples, Sec.~\ref{sec:ssh} where the extended state appear when the fixed points break the underlying Hermiticity of the model, and in Sec.~\ref{sec:ladder}, where this behaviour occurs by breaking a PT symmetry.
For general nonlinear maps, such manifolds are complicated global features, and therefore much more complex structures than the fixed points that determine the interface states described in this work.
This then provides a natural point of contact to approaches based on analytical continuation of the band structure to complex propagation factors  or generalized boundary conditions \cite{Imu19,Yokomizo19}.

Overall, these observations suggest that the approach presented in this work not only serves as a comprehensive framework of the analysis and design of interface states in a large class of settings, but also presents a starting point to consider a wide range of further aspects of these systems.

\begin{acknowledgments}
This work was funded by EPSRC via Program Grant No. EP/N031776/1, and by the Japanese Society for the Promotion of Science via JSPS invitational Fellowship No. L20556. All data generated for this work has been directly
processed into the figures.
\end{acknowledgments}

\appendix

\section{Symmetry constraints}
\label{app:sym}
The renormalization approach presented in this paper does not rely on symmetries, but nonetheless allows us to take these efficiently into account. This is due to systematic constraints and dualities of the renormalization map \eqref{eq:genmap}, which we here specify concretely for particularly relevant cases.  For compactness, we denote the renormalization map as
\begin{equation}
M(X;E)=Y,
\end{equation}
thereby keeping track of its dependence on the generally complex energy. Given a symmetry of the underlying microscopic model, we can then formulate dualities that relate maps at transformed energies and initial conditions, constraints on the fixed points of the map at symmetry-protected energies, and constraints on the flow itself if the initial conditions also preserve this symmetry.

We start from the constraints arising from Hermiticity, corresponding to systems with
 $H=H^\dagger$, $W=V^\dagger$. The renormalization map then obeys the duality
\begin{equation}
M(X^\dagger ;E^*)=Y^\dagger.
\end{equation}
It follows that at real energy $E$, any fixed point is either Hermitian, or has a Hermitian-conjugated partner. If furthermore the initial conditions are Hermitian, the flow is constrained to remain Hermitian.

For a (possibly non-Hermitian) system with conventional time reversal, $H=H^*$, $W=W^*$, $V=V^*$, the renormalization map obeys the duality
\begin{equation}
M(X^* ;E^*)=Y^*.
\end{equation}
At real energy $E$, fixed points are real or appear in complex-conjugated pairs. If furthermore the initial conditions are real, this feature is preserved under the flow.

In a reciprocal system, with $H=H^T$, $W=V^T$, the renormalization map obeys
\begin{equation}
M(X^T;E)=Y^T.
\end{equation}
At any energy $E$, fixed points are symmetric under transposition or appear in mutually transposed pairs. For transposition-symmetric initial conditions, this feature is then preserved under the flow.

Analogous constraints also arise from generalized symmetries appearing in Hermitian and non-Hermitian topological classifications \cite{Has10,Qi11,Beenakker2015,Kaw19,Oza19,zhou2019,ashida2020,Ota20,Okuma2022}.
For a chiral symmetry $ZHZ=-H$, $ZVZ=-V$, $ZWZ=-W$ induced by a unitary involution $Z$, the duality relation of the map takes the form
\begin{equation}
M(-ZXZ;-E)=-ZYZ.
\end{equation}
For a particle-hole symmetry, where analogously
$ZHZ=-H^*$, $ZVZ=-V^*$, $ZWZ=-W^*$, the duality relation of the map takes the form \begin{equation}
M(-ZX^*Z;-E^*)=-Z Y^*Z.
\end{equation}
The constraints on the fixed points or flow then appear when $E=0$ (chiral symmetry) or $E$ is purely imaginary (particle-hole symmetry).
Furthermore, for non-Hermitian systems, we can separately consider the then independent variant
$ZHZ=-H^\dagger$, $ZVZ=-W^\dagger$, where
\begin{equation}
M(-ZX^\dagger Z;-E^*)=-ZY^\dagger Z
\end{equation}
and the constraints appear for purely imaginary $E$, as well as the variant $ZHZ=-H^T$, $ZVZ=-W^T$,  where
\begin{equation}
M(-ZX^T Z;-E)=-ZY^T Z
\end{equation}
and the constraints appear at $E=0$.

Because of its conceptual importance, we also mention the case of a PT symmetry, as realized by
$PHP=H^*$, $PVP=W^*$  with a unitary involution $P$ that physically encodes a reflection of the system. This is a non-symmorphic symmetry (a symmetry that does not commute with the translation operator), and therefore is broken by edges or interfaces, unless these are placed in specific, symmetry-preserving arrangements.
In keeping with this, we then find the relation
\begin{equation}
M(PX^*P;E^*)=P[H+V(E-X)^{-1}W]^*P,
\end{equation}
where the term in brackets on the right-hand side corresponds to the renormalization map starting from the other end of the system.
Direct constraints only apply when we impose further conditions. These naturally arise, e.g., for a PT-symmetric interface in the middle of a system, as in the pristine configuration of the model in Sec.~\ref{sec:ladder}. In a parity-invariant basis, the constraints are then of the same form as in a system with time-reversal symmetry.

\begin{figure}[t]
\includegraphics[width=0.7\linewidth]{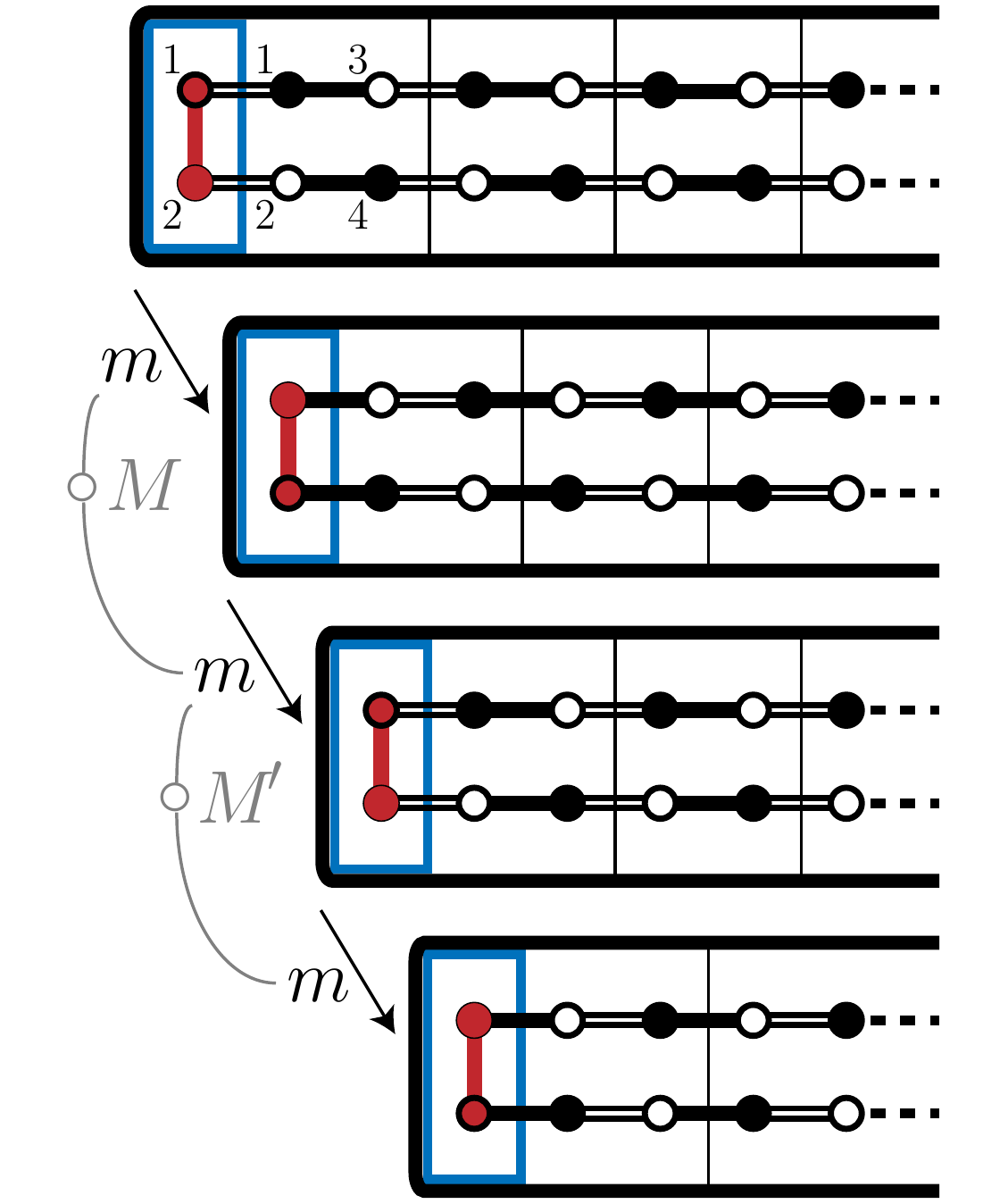}
\caption{Origin of the decomposition rule \eqref{eq:mapdimer} of the renormalization map for the dimer chain, as explained in App.~\ref{sec:appdimer}.}
\label{fig:dimeralt}
\end{figure}

\section{Alternative derivation of the renormalization map in the dimer chain}
\label{sec:appdimer}
In the main text, we derived the renormalization map of the dimer chain of Sec.~\ref{sec:dimer} in a way so it that directly acts in the interface parameter space.
Here, we describe an alternative derivation that highlights the inherent flexibility of the general approach, and illuminates the origin of the composition rule \eqref{eq:mapdimer}.

As shown in Fig.~\ref{fig:dimeralt}, this is achieved by reducing the interface to a single dimer, which exploits the short-ranged nature of the couplings. As also indicated there, it is then convenient to redefine the unit cell so that the roles of the couplings $s$ and $t$ and the onsite potentials $\pm u_0$ in the bulk are effectively interchanged, and to relabel the sites inside this cell to mimic the interface cell.
We then set the defect data to
\begin{align}
h=\begin{pmatrix}
    u_L & d \\
    d & u_R
  \end{pmatrix},
  \quad v=w^T=
  \begin{pmatrix}
    s & 0 & 0 & 0 \\
    0 & s & 0 & 0
  \end{pmatrix},
\end{align}
and the bulk data to
\begin{align}
H=\begin{pmatrix}
    -u_0 & 0 &  t & 0 \\
    0 & u_0  &  0 & t\\
    t & 0 & u_0 & 0 \\
    0 & t & 0 & -u_0 \\
  \end{pmatrix},
  \quad V=W^T=
  \begin{pmatrix}
    0 & 0 & 0 & 0 \\
    0 & 0 & 0 & 0 \\
    s & 0 & 0 & 0 \\
    0 & s & 0 & 0
  \end{pmatrix}.
\end{align}
The preparatory step \eqref{eq:preproc} generates the renormalized Hamiltonian
\begin{equation}
H^{(1)}=\begin{pmatrix}
    u_L' & d' &  t & 0 \\
    d' & u_R'  &  0 & t\\
    t & 0 & u_0 &  0\\
    0 & t & 0 & -u_0 \\
  \end{pmatrix},
  \label{eq:dimerh1alt}
\end{equation}
with the primed parameters given by the map $m_{u_0,s}$, as specified in Eq.~\eqref{eq:fullmap}.
The iteration \eqref{eq:genmap} then generates a renormalized Hamiltonian $H^{(n)}$ of the same structure as
\eqref{eq:dimerh1alt}, but with the primed parameters transformed by two further iterations of Eq.~\eqref{eq:fullmap}, where
the first iteration takes the bulk parameters $(-u_0,t)$ and the second the parameters $(u_0,s)$.
This then constitutes the renormalization map, which now takes the form
\begin{equation}
M'=m_{u_0,s}\circ m_{-u_0,t}.
\label{eq:mapdimer2}
\end{equation}
This differs from the map \eqref{eq:mapdimer} in the order of the applications. We see that each substep $m_{u,r}$ corresponds to absorbing a pair of bulk sites into the interface, instead of absorbing a pair of dimers as done in the main text.
The two descriptions are completely equivalent, as results can be translated into the space of bare interface parameters by $m_{u_0,s}$.

%\bibliography{refs}


\end{document}